\documentclass[conference]{IEEEtran}

\ifCLASSINFOpdf
\else
\fi

\usepackage[normalem]{ulem}
\usepackage{wrapfig}
\usepackage{lipsum}

\usepackage[parfill]{parskip}
\usepackage{booktabs} 
\usepackage{float}
\usepackage{stfloats}

\usepackage{graphicx}
\usepackage{epstopdf}
\usepackage{epsfig}
\usepackage[font=small,subrefformat=parens]{subcaption}
\usepackage{lipsum}
\usepackage{multirow,multicol,array}
\usepackage{tabularx}
\usepackage{pifont}
\usepackage{setspace}
\usepackage{mwe}
\usepackage{enumitem}
\usepackage{color,soul}
\usepackage{algorithm,algpseudocode,amsmath}

\usepackage{array}
\newcolumntype{P}[1]{>{\centering\arraybackslash}p{#1}}
\newcolumntype{M}[1]{>{\centering\arraybackslash}m{#1}}
\DeclareCaptionLabelFormat{andtable}{#1~#2  \&  \tablename~\thetable}
\DeclareCaptionLabelFormat{andfigure}{\tablename~\thetable \ \&  #1~#2}

\algnewcommand\algorithmicforeach{\textbf{for each}}
\algdef{S}[FOR]{ForEach}[1]{\algorithmicforeach\ #1\ \algorithmicdo}

\makeatletter
\def\blfootnote{\gdef\@thefnmark{}\@footnotetext}
\makeatother

\makeatletter
\def\footnoterule{\kern-3\p@
  \hrule \@width 2in \kern 2.6\p@} 
\makeatother

\newlength{\bibitemsep}
\newlength{\bibparskip}
\let\oldthebibliography\thebibliography
\renewcommand\thebibliography[1]{%
  \oldthebibliography{#1}%
  \setlength{\parskip}{\bibitemsep}%
  \setlength{\itemsep}{\bibparskip}%
}

\newcommand{\mycomment}[1]{}
\newcommand{\ignore}[1]{}

\usepackage{color}
\usepackage{soul}
\usepackage{xspace}

\begin{document}
\title{TraceTracker: Hardware/Software Co-Evaluation for Large-Scale I/O Workload Reconstruction}

\author{\IEEEauthorblockN{
Miryeong Kwon\IEEEauthorrefmark{1},
Jie Zhang\IEEEauthorrefmark{1},
Gyuyoung Park\IEEEauthorrefmark{1},
Wonil Choi\IEEEauthorrefmark{2},\\
David Donofrio\IEEEauthorrefmark{3},
John Shalf\IEEEauthorrefmark{3},
Mahmut Kandemir\IEEEauthorrefmark{2}
and Myoungsoo Jung\IEEEauthorrefmark{1}}
\IEEEauthorblockA{
Computer Architecture and Memory Systems Laboratory, School of Integrated Technology, Yonsei University\IEEEauthorrefmark{1},\\
Pennsylvania State University\IEEEauthorrefmark{2},
Lawrence Berkeley National Laboratory\IEEEauthorrefmark{3}\\
\IEEEauthorrefmark{1}mkwon@camelab.org,
\IEEEauthorrefmark{1}jie@camelab.org,
\IEEEauthorrefmark{1}gyuyoung@camelab.org,
\IEEEauthorrefmark{2}wuc138@cse.psu.edu, \\
\IEEEauthorrefmark{3}ddonofrio@lbl.gov,
\IEEEauthorrefmark{3}jshalf@lbl.gov,
\IEEEauthorrefmark{2}kandemir@cse.psu.edu,
\IEEEauthorrefmark{1}mj@camelab.org}}
\maketitle

\begin{abstract}
Block traces are widely used for system studies, model verifications, and design analyses in both industry and academia. While such traces include detailed block access patterns, existing trace-driven research unfortunately often fails to find true-north due to a lack of runtime contexts such as user idle periods and system delays, which are fundamentally linked to the characteristics of target storage hardware. In this work, we propose \emph{TraceTracker}, a novel hardware/software co-evaluation method that allows users to reuse a broad range of the existing block traces by keeping most their execution contexts and user scenarios while adjusting them with new system information. Specifically, our TraceTracker's software evaluation model can infer CPU burst times and user idle periods from old storage traces, whereas its hardware evaluation method remasters the storage traces by interoperating the inferred time information, and updates all inter-arrival times by making them aware of the target storage system. We apply the proposed co-evaluation model to 577 traces, which were collected by servers from different institutions and locations a decade ago, and revive the traces on a high-performance flash-based storage array.
The evaluation results reveal that the accuracy of the execution contexts reconstructed by TraceTracker is on average 99\% and 96\% with regard to the frequency of idle operations and the total idle periods, respectively. 
\end{abstract}

\let\thefootnote\relax\footnote{This paper is accepted by and will be published at 2017 IEEE International Symposium on Workload Characterization. This document is presented to ensure timely dissemination of scholarly and technical work.}

\section{Introduction}
\label{sec:intro}
Tracing block accesses is a long-established method to extract and tabulate various system parameters. 
A set of collected I/O instructions, referred to as a block trace, can provide valuable insights into design tradeoffs and can be used for the implementations of various software subsystems and hardware components in storage stacks. Therefore, many proposals utilize a wide spectrum of block traces for system characterizations, model verifications, and design analyses \cite{kim2010workload,mohan2010learned,narayanan2008everest}. 
Nevertheless, it is non-trivial, and ever-challenging to appropriately record block accesses on various large servers. Thus, open-license block traces, collected on different institutions and server locations, are extensively used in the computer and system communities \cite{verma2010srcmap, kavalanekar2008characterization, narayanan2008write, koller2010deduplication}.

While these traces include detailed block access information, they can also lead to wrong results and conclusions for some simulation-based analyses and design studies. Specifically, time information (i.e., inter-arrival time) on traces is intrinsically connected to the performance characteristics of the target storage. Since modern storage systems are undergoing significant technology shifts, different performance exhibited by new hardware can result in different I/O timing and user application behaviors. Furthermore, open-license block traces are collected on old systems that employ many hard disk drives (HDDs) designed a decade ago, which in turn can make system analysis and evaluation based on such block traces significantly different from the actual results that reflect the real characteristics of modern systems.

Even though the limited timing information is a matter for system research, it is extremely challenging to collect comprehensive information on a variety of servers and large-scale computing systems by incorporating many important (but unpredictable) user scenarios. For example, Microsoft's exchange server workloads \cite{kavalanekar2008characterization}, which are one of most popular block traces in system community, recorded detailed I/O patterns across multiple production clusters, which were generated by 5,000 users. Even if one tries to retrace the workloads by constructing such servers with modern storage like solid state drives (SSDs), it is difficult to capture all system delays, idle operations and non-deterministic timing behaviors generated by thousands of users. To address these challenges, some replay methods statically accelerate the old traces to study peak performance \cite{trushkowsky2011scads,jeong2014lifetime,weddle2007paraid}. However, these overly-simplified ``Acceleration'' methods are too imprecise to remaster the time information of the workloads. 
There also exist dynamic approaches that revise the block traces by issuing actual I/Os to a real system \cite{mesnier2007trace,zhu2005tbbt,chen2011tpc}. These ``Revision'' methods can make the inter-arrival times of workloads more realistic, but they can also lose other important runtime contexts such as user idle periods and system delays.

\begin{figure}
\centering
\includegraphics[width=1\columnwidth]{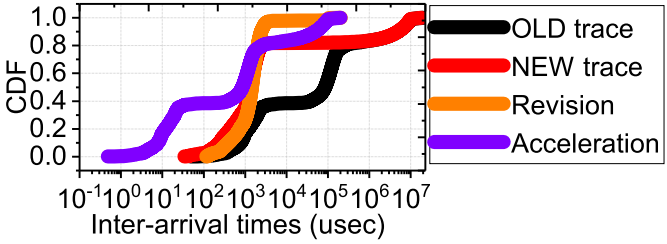}
\caption{Cumulative distribution function (CDF) for inter-arrival times observed by different methods and systems.\vspace{-8pt}}
\label{fig:intro_motiv}
\end{figure}

To be precise, we evaluate the inter-arrival times generated by an acceleration method (\texttt{Acceleration} \cite{jeong2014lifetime}) and a revision method (\texttt{Revision} \cite{chen2011tpc}), and the results are shown in Figure \ref{fig:intro_motiv} in the form of a cumulative distribution function (CDF). In this evaluation, we generated 70 million instructions whose I/O patterns are same as a Microsoft's network storage file server \cite{kavalanekar2008characterization}, and are issued to an HDD-based system node (\texttt{OLD}) and a SSD-based system node (\texttt{NEW}), respectively. Specifically, 14 million instructions are issued in an asynchronous fashion, and we injected user idle operations that account for 20\% of the total instructions to make I/O access more realistic. The same patterns are collected from both \texttt{OLD} and \texttt{NEW} for a fair comparison. The traces on \texttt{OLD} are used for \texttt{Acceleration} while \texttt{Revision} is implemented by reconstructing the workloads by replaying them on the SSD-based storage node. As shown in the figure, the first half of the distribution curve of \texttt{Acceleration} exhibits shorter inter-arrival times than that of the actual target system (\texttt{NEW}) by 88\% on average, while losing 98\% of user idle times, compared to the target system. Even though the timing trend of \texttt{Revision} appears similar to that of \texttt{NEW}, it still exhibits longer inter-arrival times at the first half of CDF curve than the ones of \texttt{NEW} by 16\%, on average. More importantly, \texttt{Revision} fails to capture 18\% of user idle operations and 69\% of total idle periods, observed in the real system, \texttt{NEW}.

In this paper, we propose \emph{TraceTracker}, a novel hardware/software co-evaluation method that allows users to reuse a broad range of the existing block traces by keeping most of their execution contexts and user behaviors while adjusting them to the new system information. Specifically, our proposed TraceTracker's software evaluation model can infer CPU burst times and user idle periods from old-fashioned block traces, whereas its hardware evaluation method remasters the block traces by interoperating the inferred time information as well as renews all inter-arrival times by being aware of the target storage system. The proposed software and hardware co-evaluation methods can be implemented by using publicly-available benchmark tools such as FIO \cite{axboe2011flexible}.

The main \textbf{contributions} can be summarized as follows: 

$\bullet$ \textit{Reviving the timing information for diverse workloads.} There are several workloads that provide no specific information or descriptions of the underlying storage trace collection environment. In this work, we analyze a diverse set of large-scale workloads and provide an inference model that estimates the relative time costs of an I/O request service. This inference model evaluates the realistic idle time that can capture the system and user behaviors from the traditional block traces by dividing the arrival time into a channel delay, a device time and an idle time. It decomposes I/O subsystem latency by analyzing the probability density functions and cumulative distribution functions of the inter-arrival times as well as being aware of the given request sizes and operation types.


$\bullet$ \textit{Inference automation and hardware/software co-evaluation.} Analyzing extensive out-of-date block traces is non-trivial, and reconstructing the traces is not a one-shot process as the target system will keep shifting its underlying storage technology. In this work, we reify the proposed inference model by automating our graph classification method and steepness analysis, each of which is used to examine massive trace data and speculate the underlying I/O system latency. With the timing information deduced by the proposed inference automation, TraceTracker simulates the old system behaviors and emulates I/O services on a real target system. TraceTracker also performs post-processing to revive asynchronous/synchronous information on new emulated traces. To verify the proposed trace reconstruction method, we introduce several verification metrics such as user idle detection and length. Even in the case of no runtime information being available on the trace collections, our TraceTracker can detect 99\% of system delays and idle periods appropriately and secure the corresponding idle periods by 96\% of a real execution, on average.

$\bullet$ \textit{Massive trace reconstruction and analysis.}
In this work, we reconstructed 577 traces\footnote{All the traces collected for this paper are available for download from http://trace.camelab.org.} that cover a diverse set of I/O workloads of large-scale computing systems, such as web services, data mining and network file system severs, and performed a comprehensive analysis of the reconstructed block traces. While previous work \cite{jeong2014lifetime} claimed that 50\% of write requests have time intervals that are 2x longer than the effective device operation latency, even after accelerating the block traces (of the same workloads that we tested) by 100x, we observed that the number of time intervals that have idle periods is less than 39\% of the total number of I/O requests.
Note that the majority of idle periods in all the block traces are found in 1 millisecond, which is also 10\% shorter than the one reported by the prior study \cite{jeong2014lifetime}.
\ignore{
** will be moved to background section.
For example, I/O off-loading techniques \cite{Everest, finding out more} exploit idle bandwidth to multiplex the storage bandwidth resources in data center. Similarly, automated load balancing and storage virtualization \cite{BASIL, hui wang (12th FAST), finding out more} interleave the live data migration with incoming I/O requests among different storage subsystems by extracting I/O latency and inter-arrival times. 
Even though all these approaches demonstrated great performance improvements by maximizing the resource utilization, the inter-arrival times and idle timing behaviors that these work leveraged were collected on out-of-date systems \cite{}. 
In addition to these system-level studies, device-level researches may also introduce the improper results, which should be revised in state-of-the-art systems. For example, modern SSDs perform time-consuming processes but functionally-critical tasks in background by utilizing storage idle times or inter-arrival times. For example, \cite{jeong2013improving, jeong2014lifetime} serve non-urgent requests by stealing system idle or slack time to improve the endurance of flash media. Similarly, \cite{changfastread} migrates data between fast and slow flash blocks using idle time, and \cite{jang2010efficient,wang2012efficient,lin2013efficient} perform block merges or garbage collections as background using a slack in inter-arrival times. While all these novel device-level optimizations are verified based on diverse real-world traces \cite{}, the inter-arrival times and idle periods of evaluated traces are by far away from the modern storage system due to significant technology shifts at storage-level.    
}

\section{Background}
\label{sec:background}
\begin{figure}
\centering
\begin{subfigure}{0.28\columnwidth}
\includegraphics[width=\columnwidth]{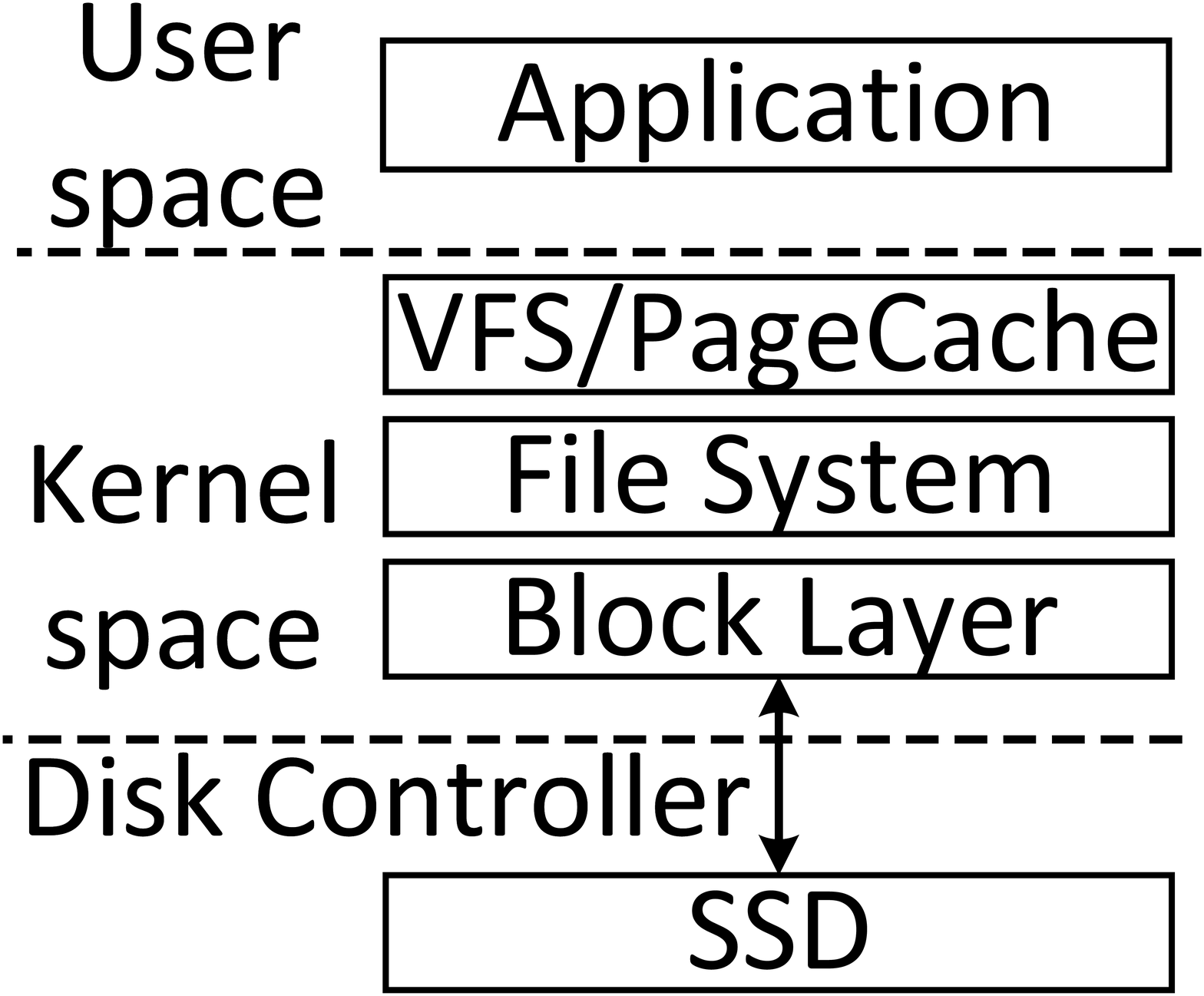}
\caption{Storage stack.}
\label{fig:back_IOStack}
\end{subfigure}
~
\begin{subfigure}{0.62\columnwidth}
\includegraphics[width=\columnwidth]{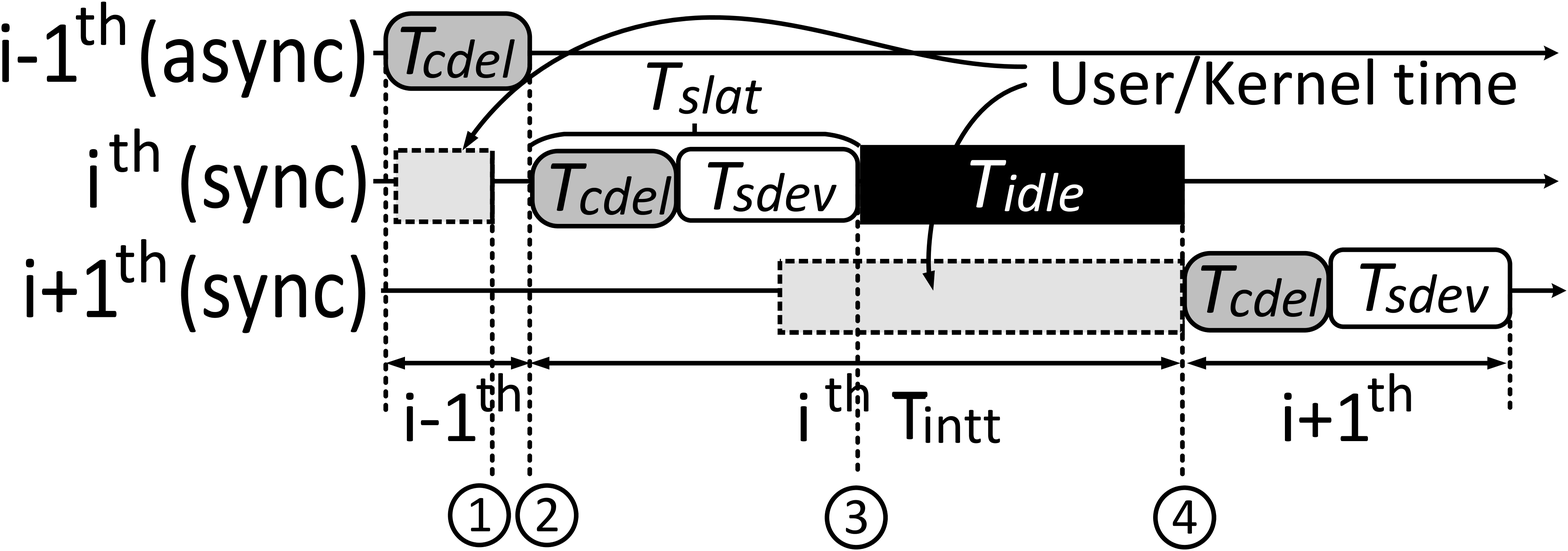}
\caption{Block request timing example.}
\label{fig:back_execute}
\end{subfigure}
\vspace{8pt}
\caption{Storage-level I/O information.\vspace{-8pt}}
\end{figure}

In this section, we first explain the storage-level I/O information from the perspective of a storage stack and block request timing sequence.  We then introduce the existing trace revision methods and discuss their limitations.

\subsection{Storage-Level I/O Information}

\noindent \textbf{Storage stack.} Figure \ref{fig:back_IOStack} illustrates a typical storage stack from an application to the underlying storage.
Once the application makes an I/O request, it is required to switch from the user mode to the kernel mode and jump into an entry-point of a virtual file system (VFS).
VFS then copies the corresponding target data from a user buffer to a kernel buffer (referred to as page cache \cite{harty1992application}) and forwards the request to the underlying file system.
During this time, the mode switch consumes CPU cycles for handling system calls and storing task states in addition to copying the buffers.
The file system then looks up the physical locations (indicated by request) and submits this information to the block layer.
Finally, the block layer partitions the translated information, including logical block address and request size (in terms of the number of sectors) into multiple packets (or transactions).
Note that, before submitting the actual information to the underlying storage, the multiple layers in the storage stack consume CPU cycles for mode switches, data copies and address translations.
Note also all open-license block traces are also typically collected underneath the block layer.
In cases where there are no system delays or application idleness, the user/kernel specific CPU bursts can overlap with storage bursts, which make the computational cycles that upper software modules consume hide behind the inter-arrival times of multiple I/O requests at the block traces.

\noindent \textbf{Block request timing.} Figure \ref{fig:back_execute} shows the timing diagram of block requests, which can be captured from underneath the block layer.
There are three requests, denoted by $(i-1)^{th}$, $i^{th}$ and $(i+1)^{th}$.
In this example, the $(i-1)^{th}$ request is issued asynchronously, whereas all other requests are issued synchronously.
Since the asynchronous block request does not need to wait for the response from the underlying device, there is only a delay caused by the storage interface (i.e., channel) data movement and corresponding data packet. This channel delay is referred to as $T_{cdel}$.
In this work, the I/O subsystem latency, called $T_{slat}$, consists of $T_{cdel}$ and the actual device time taken by the storage to service the request, denoted by $T_{sdev}$.
The $i^{th}$ request is ready to be submitted to the storage at \ding{192}, and therefore, it is issued at \ding{193}.
Even though $T_{slat}$ is finished at \ding{194}, the user/kernel consume some computation cycles, and the $(i+1)^{th}$ request is not available.
This in turn leads to $T_{idle}$  by \ding{195}. Once the $(i+1)^{th}$ request is prepared by the upper layers, it can be served with its $T_{cdel}$ and $T_{sdev}$.
Note that, in addition to this kind of system delay, $T_{idle}$ represents the time when a user or application does nothing. In this example of block request timings, the inter-arrival time, called $T_{intt}$, is defined by the time period between \ding{193} and \ding{195}.

\subsection{Trace Revision}
Even though existing block traces can cover different kinds of system configurations and various user scenarios, most publicly-available conventional block traces \cite{application2007umass,verma2010srcmap, kavalanekar2008characterization, narayanan2008write, koller2010deduplication} were collected on HDD-based storage systems around a decade ago.
Since then, however, storage systems have been dramatically changed, as modern servers start to adopt flash-based storage to boost performance and most server workloads significantly changed.
Since the block traces are intensively studied and used for demonstrating the effectiveness and performance impacts of many system research proposals \cite{zhang2013warming,jeong2014lifetime,narayanan2009migrating,soundararajan2010extending}, these traces need to be mapped to new block traces by considering the new storage system characteristics.  

Several approaches exist to reconstruct traditional block traces \cite{trushkowsky2011scads, jeong2014lifetime, weddle2007paraid, mesnier2007trace, zhu2005tbbt, chen2011tpc}.
First, the acceleration methods (\texttt{Acceleration}) \cite{trushkowsky2011scads, jeong2014lifetime, weddle2007paraid} can artificially shorten inter-arrival times to compensate for the low throughput exhibited by HDD-based storage.
However, since this method can only resize the inter-arrival times without considering the block request timings, it can remove critical information, such as $T_{cdel}$ and $T_{idle}$ from the traces. 
For example, if the average $T_{intt}$ of a workload is 50 ms and the acceleration factor is 100, the reconstructed trace exhibits 500 us for its average $T_{intt}$.
This removes the most $T_{cdel}$ and $T_{idle}$, and can even make $T_{sdev}$ unrealistic as there is no contexts for target device, system, and user behaviors.
Instead of simply accelerating inter-arrival times, there is a (\texttt{Revision}) to revise target workloads by replaying the corresponding block traces on a real system \cite{mesnier2007trace, zhu2005tbbt, chen2011tpc}.
While this would have more realistic $T_{cdel}$ and $T_{sdev}$, it cannot appropriately accommodate $T_{intt}$, which varies across all I/O instructions in the trace.

\begin{figure}
\centering
\begin{subfigure}{0.45\columnwidth}
\includegraphics[width=1\columnwidth]{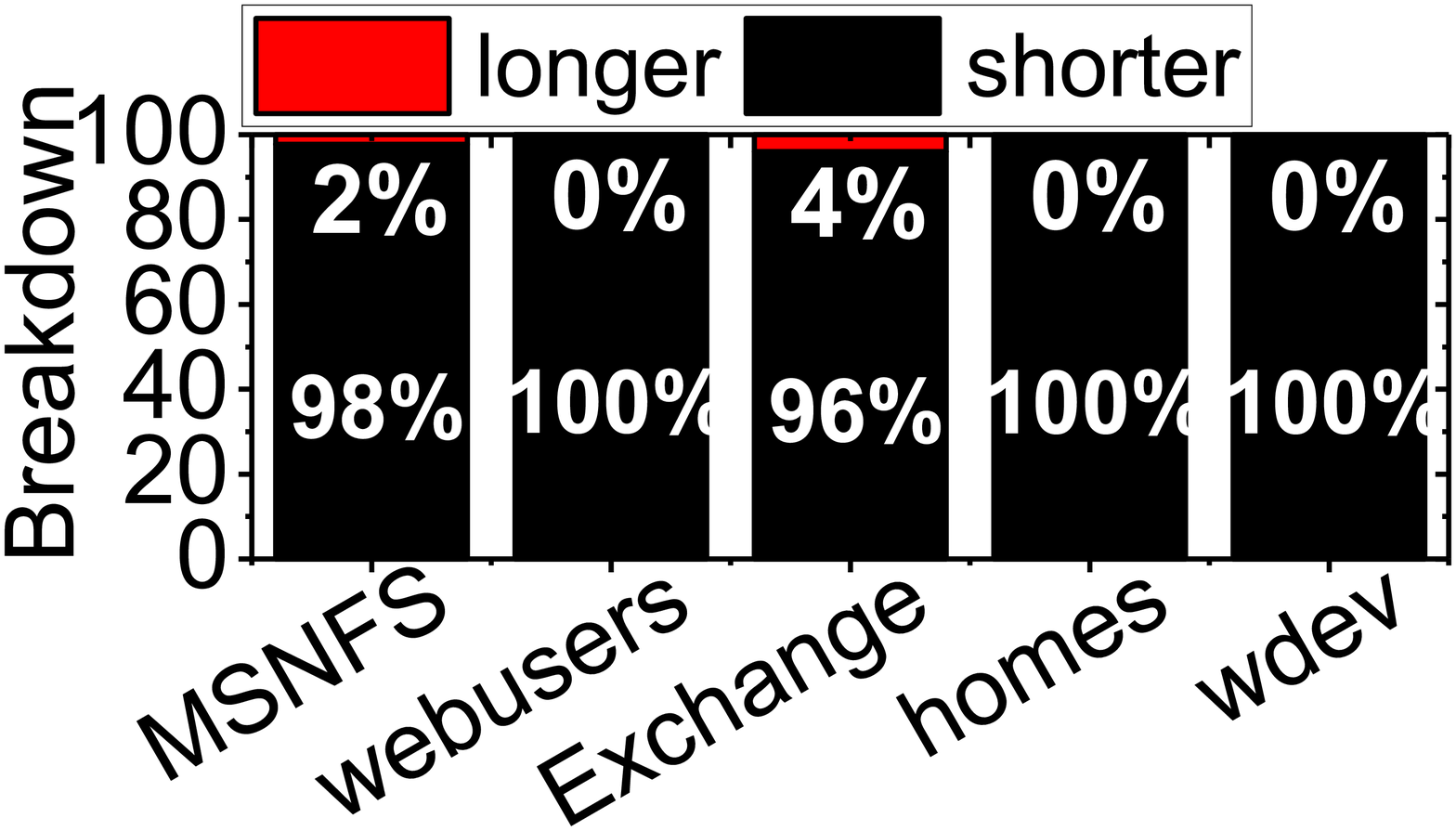}
\caption{\texttt{Acceleration}.}
\label{fig:back_accer_len}
\end{subfigure}
~
\begin{subfigure}{0.45\columnwidth}
\includegraphics[width=1\columnwidth]{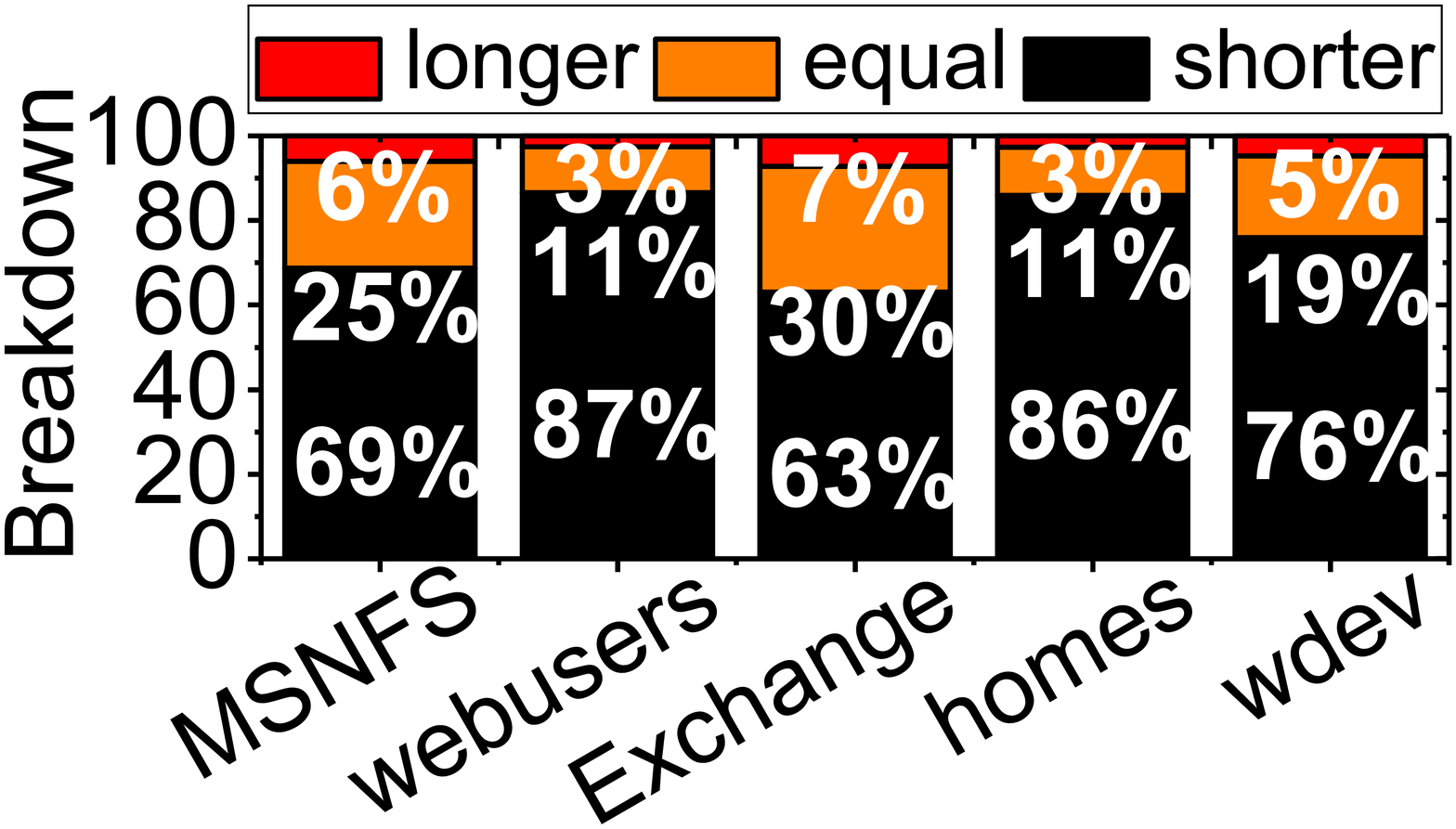}
\caption{\texttt{Revision}.}
\label{fig:back_revision_len}
\end{subfigure}
\vspace{8pt}
\caption{Differences of inter-arrival times observed by reconstructed traces and real system traces.\vspace{-10pt}}
\label{fig:back_motiv}
\end{figure}

To be precise, we also compare $T_{intt}$ observed in the SSD-based system node (\texttt{NEW}) with the ones generated by \texttt{Revision} and \texttt{Acceleration}, respectively.
The evaluation environment and scenario are the same as the test conditions described in Section \ref{sec:intro}, and \texttt{Acceleration} leverages the acceleration degree that \cite{jeong2014lifetime} uses.
We examine different $T_{intt}$ values by executing five open-license block traces (\emph{MSNFS}, \emph{webusers}, \emph{Exchange}, \emph{homes}, \emph{wdev}), which are widely used in the storage community \cite{storagesnia}, and the results are shown in Figure \ref{fig:back_motiv}.
One can observe from Figure \ref{fig:back_accer_len} that, 98.6\% of $T_{intt}$ reconstructed by \texttt{Acceleration} are shorter than actual $T_{intt}$ observed in \texttt{NEW}.
In contrast, as shown in Figure \ref{fig:back_revision_len}, $T_{intt}$ of \texttt{Revision} is on average 17.8\% accurate (i.e., `equal' in the figure).
However, most of them (77.8\% of total $T_{intt}$, on average) are shorter than the actual $T_{intt}$, which means it loses important system delays and user idle periods, $T_{intt}$.
Note that it also exhibits a $T_{intt}$, that is on average 4.3\% longer than the actual $T_{intt}$.
This is because replaying traces drops the mode contexts (i.e., asynchronous/synchronous), which fails to capture the block request timing described by the $(i-1)^{th}$ request (as shown in Figure \ref{fig:back_execute}).
Since it is difficult to capture all system delays, idle operations and non-deterministic user behaviors (there is no block trace that offers all such information to the best of our knowledge), block trace reconstruction with limited information is non-trivial and challenging work.

\section{Timing Inference for I/O Subsystems}
\label{sec:method}
\begin{figure*}
\centering
\includegraphics[width=2\columnwidth]{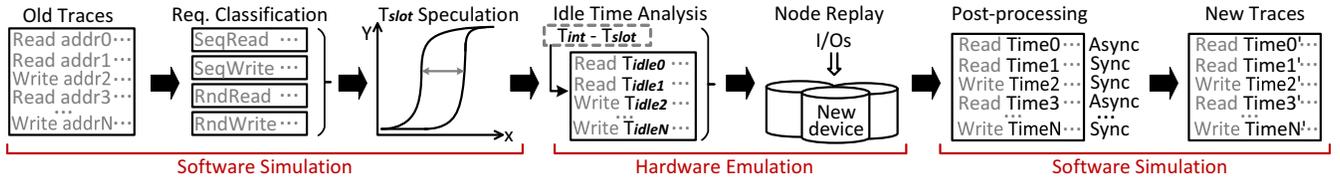}
\vspace{-8pt}
\caption{High-level view of TraceTracker.}
\label{fig:method_highlevel}
\end{figure*}

\begin{figure}
\centering
\begin{subfigure}{0.30\columnwidth}
\includegraphics[width=\columnwidth]{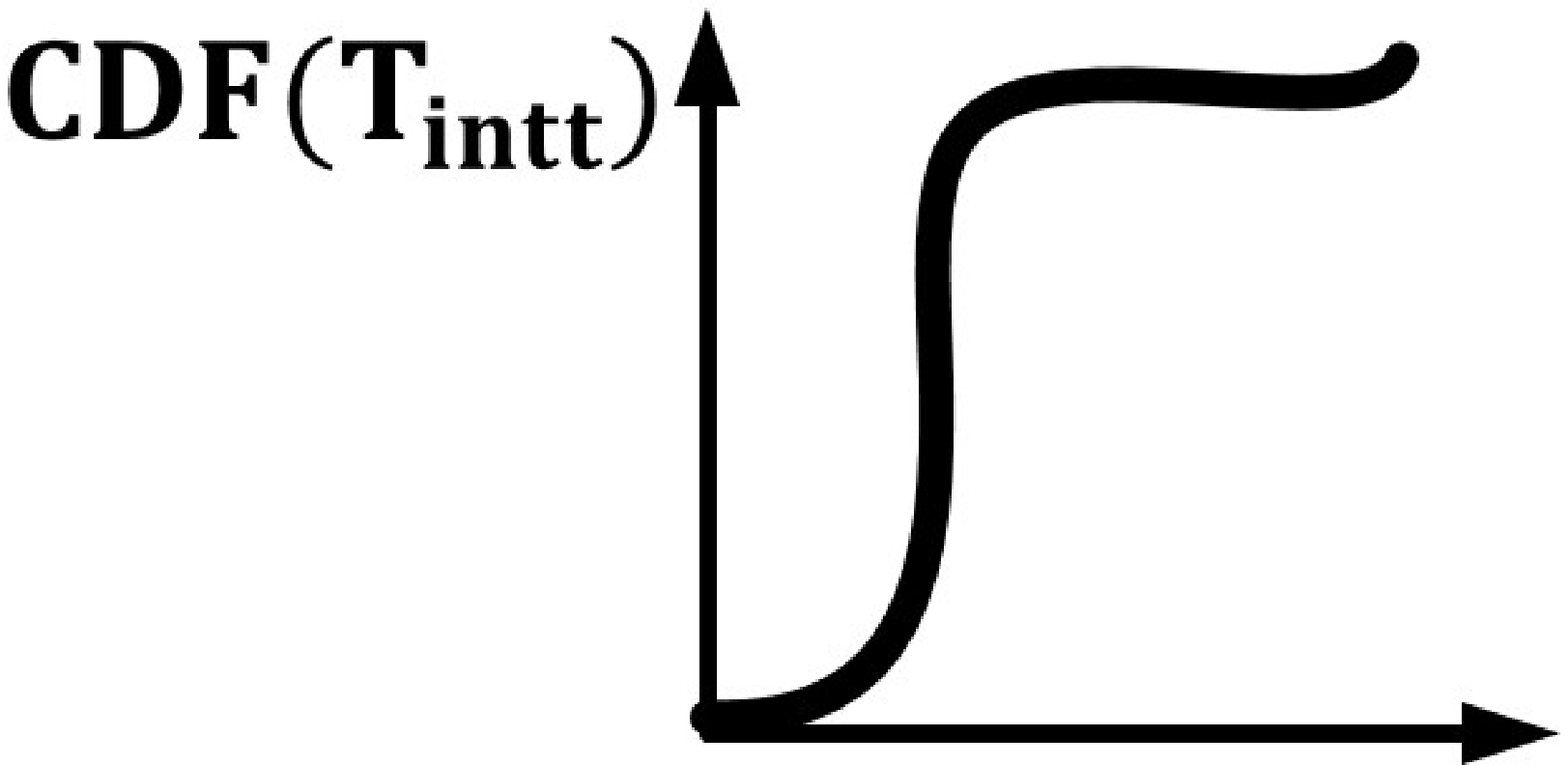}
\caption{Global maxima.}
\label{fig:eval_CDF1}
\end{subfigure}
~
\begin{subfigure}{0.30\columnwidth}
\includegraphics[width=\columnwidth]{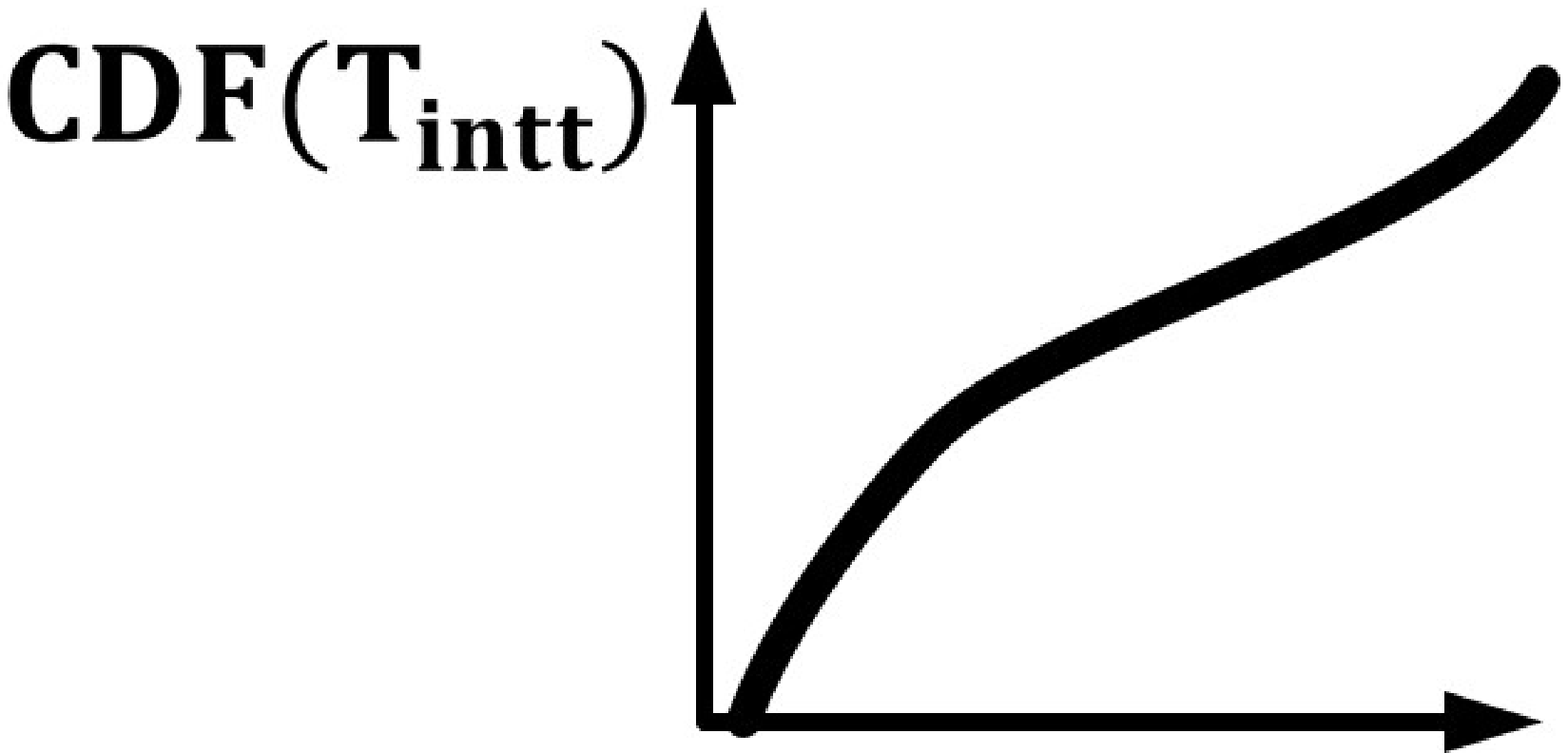}
\caption{Chunky middle.}
\label{fig:eval_CDF2}
\end{subfigure}
~
\begin{subfigure}{0.30\columnwidth}
\includegraphics[width=\columnwidth]{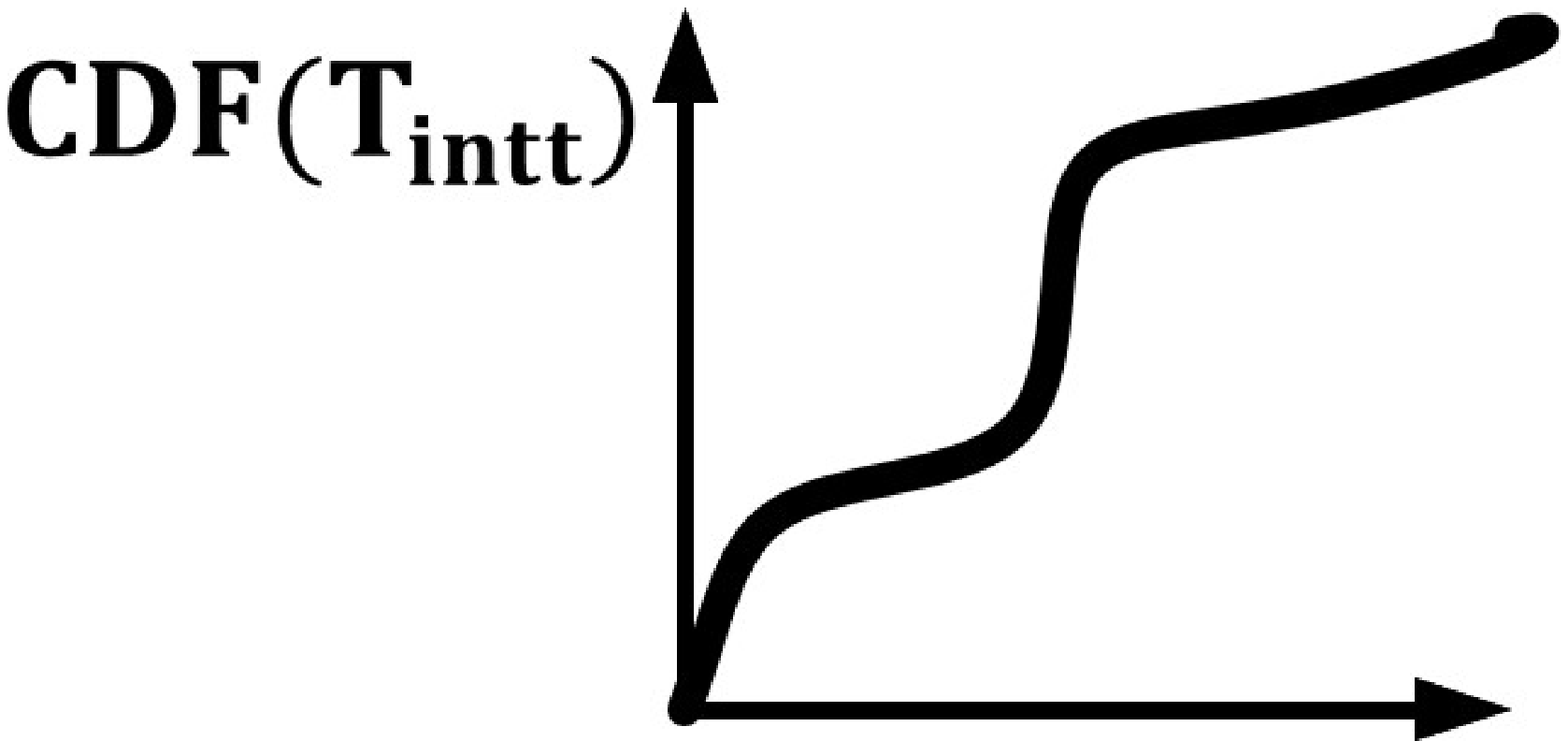}
\caption{Multi maxima.}
\label{fig:eval_CDF3}
\end{subfigure}
\vspace{8pt}
\caption{Types of CDF distribution.\vspace{-5pt}}
\label{fig:method_categorization}
\end{figure}

\noindent \textbf{Overview of TraceTacker.}
For individual I/O instructions, it is non-trivial to extract the idle time (i.e., $T_{idle}$) from old block traces since $T_{idle}$ is affected by multiple unknown system parameters and indeterministic user behaviors at the time of trace collection. 
Even though the old block traces have no runtime information, including the user behaviors, $T_{idle}$ can be inferred if we can estimate I/O subsystem latency (i.e., $T_{slat}$), which is composed of the channel delay (i.e., $T_{cdel}$) and storage device time (i.e., $T_{sdev}$). Generally speaking, $T_{idle}$ can be simply obtained by subtracting $T_{slat}$ from $T_{intt}$. Estimating $T_{slat}$ would be relatively easy if most inter-arrival times (i.e., $T_{intt}$) are similar to each other, which can make the graph, that represents the CDF of $T_{intt}$ (i.e., $CDF(T_{intt})$), steeper. As shown in Figure \ref{fig:eval_CDF1}, the graph rapidly rises in the middle, which exhibits a single maximum on its derivative.
Because almost the entire range of $CDF(T_{intt})$ is in the middle of domain and not affected by its tail, $T_{intt}$ at the global maximum of slope of CDF ($CDF(T_{intt})'$) can be considered as $T_{slat}$. However, there are many block traces whose $CDF(T_{intt})$ exhibits a much smoother slope on (e.g., chunky middle) and/or multiple maxima on its derivative. This is because $T_{intt}$ of each instruction is affected by a different runtime contexts on the target system, which often makes them vary significantly. Considering Figure \ref{fig:eval_CDF3} as an example, this workload exhibits at least two maxima on $CDF(T_{intt})'$, which can render such simple differential analysis difficult to predict $T_{slat}$ of the corresponding trace, appropriately.

Figure \ref{fig:method_highlevel} summarizes the operation of our proposed TraceTracker. In this work, as shown in the left software simulation of Figure \ref{fig:method_highlevel}, we classify all the I/O instructions traced by a workload into multiple groups based on the request size and operation type. For the multiple groups, we create multiple CDFs for $T_{intt}$, and estimate the relative time costs of $T_{sdev}$ and $T_{cdel}$ in block request timings by taking into account the different request sizes and types. This relative time cost estimation, in turn, enables us to individually calculate $T^i_{slat}$ for all $i$ numbers of I/O instructions, thereby extracting $T^{i}_{idle}$ from the target traditional block. Once we secure $T_{idle}$ that varies based on user and system timing behaviors, $T_{slat}$ can be re-evaluated by taking into account the target storage system. Specifically, we emulate the new system by regenerating the each request and issuing it on new storage with estimated $T_{idle}$. After the target trace emulation, we perform a simple post processing on the trace, which overrides the I/O timing behaviors for asynchronous mode operations by considering the old block trace and regenerated new trace. Further, while the inference logic of TraceTracker extracts the timing behaviors affected by non-deterministic user behaviors and unknown system parameters from the old block trace, the hardware emulation, and post processing parts mimic the system delay and user idle periods on a real system (target) to generate the new block trace. 

\begin{figure}
\centering
\vspace{-10pt}
\includegraphics[width=1\columnwidth]{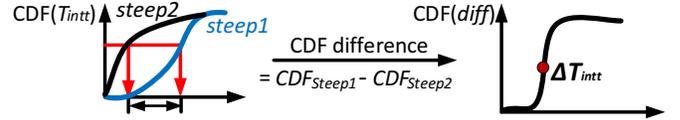}
\caption{Finding the coefficients of $T_{sdev}$ ($\beta$ or $\eta$).\vspace{-5pt}}
\label{fig:method_difference}
\end{figure}

\noindent \textbf{Inference model.}
If there is no user idle period or system delay caused by the host-side software modules, $T_{intt}$ can be similar to, or even the same as $T_{slat}$. In other words, if there is $T_{intt}$ greater than $T_{slat}$, $T_{idle}$ can be simply inferred by subtracting $T_{slat}$ from $T_{intt}$. Even though the specific information captured by $T_{slat}$ is also often not recorded and offered by the old block traces, in contrast to $T_{idle}$, it can be speculated by analyzing the distribution of $T_{intt}$. As described earlier, there is only one CDF of $T_{intt}$ if all I/O instructions in the target workload exhibit a uniformed request size. This in turn allows us to simply speculate $T_{slat}$ by referring to $T_{intt}$ at the global maxima of $CDF(T_{intt})'$. 
In cases where the target workload exhibits a wide spectrum of request sizes and types, we speculate $T_{slat}$ by inferring $T_{sdev}$ and $T_{cdel}$, separately. Since $T_{sdev}$ mainly depends on the underlying storage performance, we assume that $T_{sdev}$ follows a linear model for the sequential accesses; $T_{sdev}$ is inferred by $\beta * r_{size}$ if the type of the request is a read; otherwise, it is speculated by $\eta * r_{size}$. $\beta$ and $\eta$ are coefficient values, which will be explained shortly, and $r_{size}$ denotes the size of a request. On the other hand, $T_{sdev}$ on random accesses can be slightly longer than that of sequential accesses as it has a moving delay time, referred to as $T_{movd}$. $T_{movd}$ typically captures the seek time and rotational latency of the underlying disk \cite{ruemmler1994introduction}. 

\begin{figure}
\centering
\begin{subfigure}{1\columnwidth}
\includegraphics[width=\columnwidth]{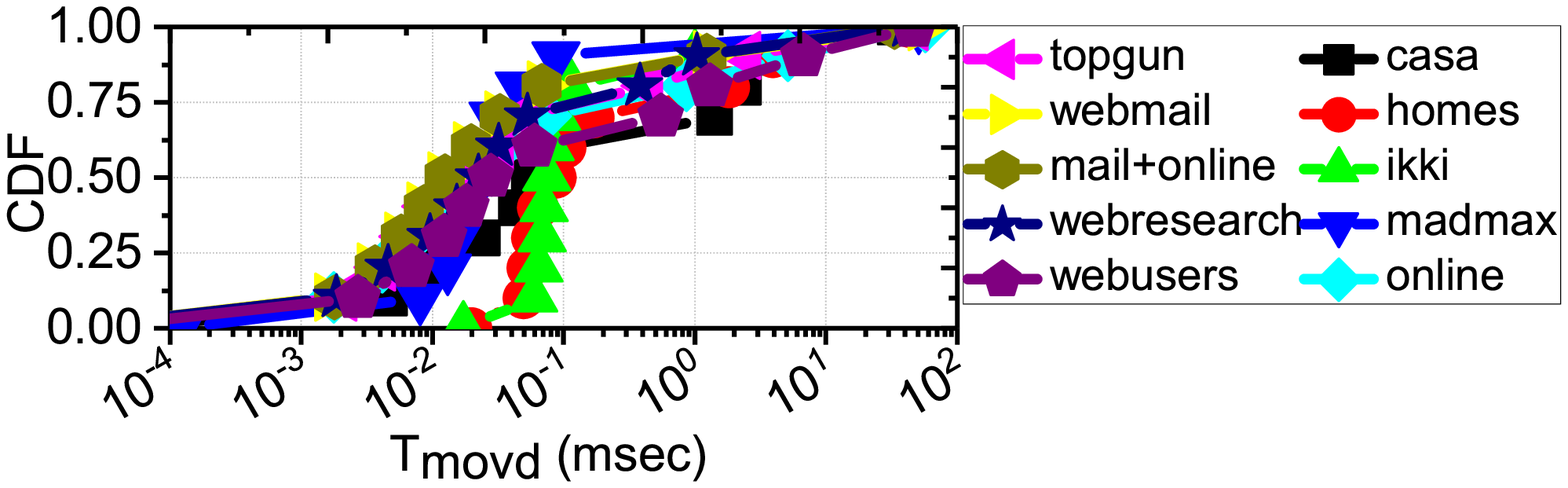}
\caption{CDF distribution of $T_{movd}$.\vspace{10pt}}
\label{fig:method_Tmovd}
\end{subfigure}
\\
\begin{subfigure}{1\columnwidth}
\includegraphics[width=\columnwidth]{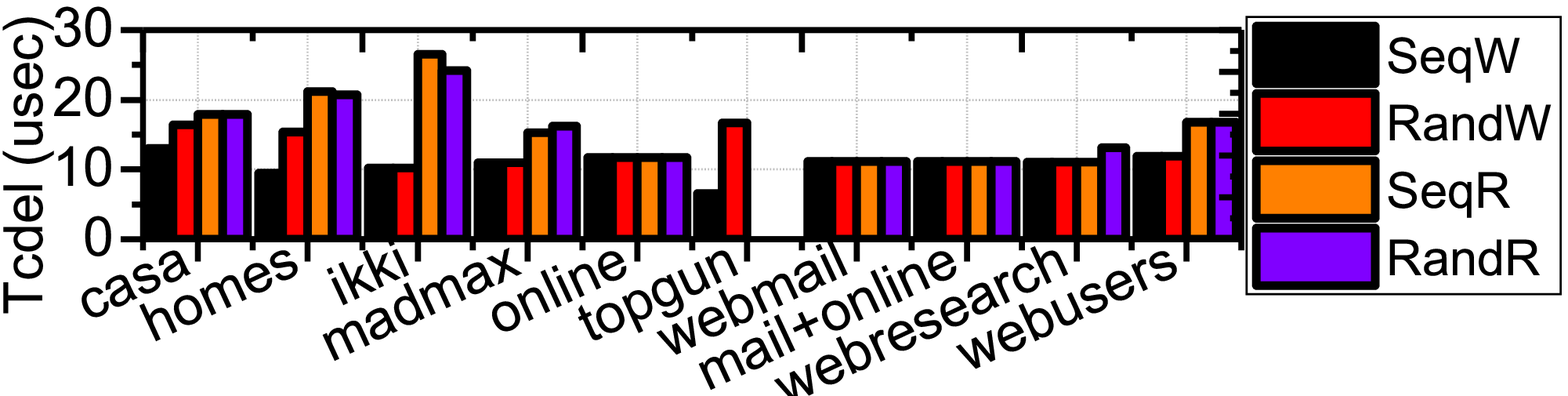}
\caption{Average period of $T_{cdel}$.}
\label{fig:method_Tcdel}
\end{subfigure}
\vspace{8pt}
\caption{The time components of $T_{slat}$ (FIU).}
\end{figure}

To model $T_{movd}$, we replay ten FIU workloads \cite{verma2010srcmap, koller2010deduplication} on an enterprise disk \cite{WDBlue} and measure $T_{movd}$ by calculating the difference between the $T^{real}_{sdev}$ and $T^{linear}_{sdev}$, each of which is $T_{sdev}$ observed on the real disk by executing random I/O accesses and generated by our linear models with sequential I/O accesses, respectively. We consider the difference between $T^{real}_{sdev}$ and $T^{linear}_{sdev}$ as $T_{movd}$, and the $CDF(T_{movd})$ results (for each workload) are plotted in Figure \ref{fig:method_Tmovd}.
As shown in the figure, each CDF exhibits a similar magnitude of gradient change with transition of $T_{movd}$. Motivated by this, we use $T_{movd}$ at the maximum of $CDF(T_{movd})'$ as the representative of the difference between $T^{real}_{sdev}$ and $T^{linear}_{sdev}$, which is referred to as $T^{rep}_{movd}$. Consequently, in this work, $T_{sdev}$ for random reads and writes can be expressed by $\beta * r_{size} + T^{rep}_{movd}$ and $\eta * r_{size} + T^{rep}_{movd}$, respectively. Using this inference model, we speculate $T_{sdev}$, which in turn allow us to infer $T_{slat}$ and $T_{idle}$. 
The specific estimation methods for each relative costs in block request timings are described below.


\noindent \textbf{Decomposition of I/O subsystem latency.}
For each workload, the coefficients of $T_{sdev}$ in our inference model, $\beta$ and $\eta$, can be estimated by using the following disintegration analysis. First, we group all I/O instructions of the workload to reconstruct into three different categories based on i) sequentiality (e.g., sequential vs. random), ii) operation type (e.g., read vs. write) and iii) request size (in terms of sectors). We then create multiple graphs of $CDF(T_{intt})$ for each request size observed in each read or write with the sequential access pattern. The proposed inference model then examines the global maxima of $CDF(T_{intt})'$ for each CDF. 
Thus, there can be $n$ maxima of $CDF(T_{intt})'$, where $n$ is the number of different I/O request sizes observed in a target workload. It then chooses the two steepest graphs of CDF, which have the two highest magnitudes of $T_{intt}$ changes among the maximas. Let us denote each of the steepest functions as $CDF_{steep1}(T_{intt})$ and $CDF_{steep2}(T_{intt})$, where $CDF_{steep1}(T_{intt})'$ is greater than $CDF_{steep2}(T_{intt})'$. As shown in Figure \ref{fig:method_difference}, we can drive $CDF(diff)$ which is CDF difference of $CDF_{steep1}(T_{intt})$ and $CDF_{steep2}(T_{intt})$ for reads and writes separately, and calculate the maximum at $CDF(diff)'$ for each.

We can then obtain the representative inter-arrival time at the maximum of $CDF(diff)'$ for reads and writes, which are referred to as $\Delta T^{read}_{intt}$ and $\Delta T^{write}_{intt}$, respectively. Let us denote the two request sizes, which are used for creating $CDF_{steep1}(T_{intt})$ and $CDF_{steep2}(T_{intt})$, as $size_{r1}$ and $size_{r2}$, respectively. We can estimate the $\beta$ and $\eta$ by calculating $\Delta T^{read}_{intt} / |size_{r1} - size_{r2}|$ and $\Delta T^{wrte}_{intt} / |size_{r1} - size_{r2}|$, respectively. Let us denote the inter-arrival times at the maximum of $CDF_{steep1}(T_{intt})'$ for reads and writes as $T'^{read}_{intt}$ and $T'^{write}_{intt}$, respectively. $T'^{read}_{intt}$ and $T'^{write}_{intt}$ are the best values that explain $T_{slat}$ of the target workload since they exclude the most of the timing effects caused by system delays and user idle periods. Next, we can obtain $T^{read}_{cdel}$ and $T^{write}_{cdel}$ by calculating $T'^{read}_{intt}  - \beta * size_{r1}$ and $T'^{write}_{intt} - \eta * size_{ref}$, where $T^{read}_{cdel}$ and $T^{write}_{cdel}$ are the channel delays for the reads and writes, respectively. 

Figure \ref{fig:method_Tcdel} shows the actual $T_{cdel}$ of the FIU workloads that we observed on the disk for each access pattern. One can observe from this figure that, while the difference of $T_{cdel}$ between reads and writes exist to some extent (e.g., \emph{ikki} and \emph{maxmax}), that of $T_{cdel}$ between random and sequential access patterns is not significant (less than 8\% and 6\%, respectively). Noting that the difference of $T_{cdel}$ is less than $T_{sdev}$ by many orders of magnitude; we believe that estimating the channel delay based the operation type of I/O requests is reasonable. 

Lastly, to estimate the relative time cost of $T_{movd}$, we also need to find $CDF^{rand}_{steep}(T_{intt})$ that has the highest magnitude of gradient change with transition of $T_{intt}$ among multiple CDFs in the group of random accesses, and estimate the inter-arrival time, $T^{rand}_{intt}$, at the maximum of $CDF^{rand}_{steep}(T_{intt})'$. Then, $T_{movd}$ can be simply calculated by subtracting $T^{read}_{cdel} + \beta * size_{ref}$ (or $T^{write}_{cdel} + \eta * size_{ref}$) from the estimated $T^{rand}_{intt}$.

\section{Implementation for Inference Automation}
\label{sec:auto}
Analyzing multiple CDF graphs is important to reconstruct old traces. While categorizing requests based on their types and sizes can be easily automated, the autonomous analysis of CDFs is non-trivial due to their discreteness. In this section, we will detail the implementation of our proposed inference model and explain how to emulate traces with inferred system delays and idle periods (i.e., $T_{intt}$) to reconstruct the traces.

\begin{figure}
\centering
\includegraphics[width=1\columnwidth]{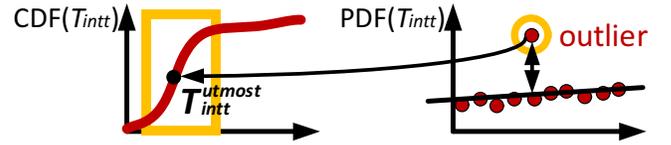}
\caption{Checking steepness of CDF distribution.\vspace{-6pt}}
\label{fig:method_max}
\end{figure}

\noindent \textbf{Graph classification.}
Since each block trace can exhibit multiple CDFs that are examined by the proposed inference model, it is time consuming to detect the two steepest graphs of $CDF(T_{intt})$, namely, $CDFsteep1(T_{intt})$ and $CDFsteep2(T_{intt})$, among them. When examining the graph, it would be possible to have a higher degree of the polynomial equation to represent CDF in mathematical expression, which also renders the process of finding $CDFsteep1(T_{intt})$ and $CDFsteep2(T_{intt})$ for the read and write instruction set of each trace difficult.

\begin{algorithm}[]
\scriptsize
\caption{CDF steepness examination.}\label{algo:method_steepness}
\begin{algorithmic}[1]

\Statex /** Step 1: Calculate PDF of inter-arrival times ($T_{intt}$)

\ForEach{$T_i$ in $T_{intt}$}
\State $PDF(T_i)$ := num($T_i$) / num(request)
\EndFor

\Statex /** Step 2: Least Square Regression

\State slope := std($PDF(T_{intt})$) / std($T_{intt}$)
\State intercept := mean($PDF(T_{intt})$) - slope * mean($T_{intt}$)
\State f(x) := slope * x + intercept

\Statex /** Step 3: Find outliers
\State margin := var($PDF(T_{intt})$) / 2
\ForEach{$T_i$ in $T_{intt}$}
\State distance := $PDF(T_i)$ - f($T_i$)
\If {distance \textgreater margin}
\State outliers.append($PDF(T_i)$)
\EndIf
\EndFor

\Statex /** Step 4: Calculate CDF steepness
\State $T_{intt}^{utmost}$ := max(outliers)
\State steepness := distance(f($T_{intt}^{utmost}$),PDF($T_{intt}^{utmost}$))
\Statex
\end{algorithmic}
\vspace{-10pt}
\end{algorithm}

One simple but effective method to check the steepness of each graph is to analyze probability density distribution (PDF), instead of examining the derived function of $CDF(T_{intt})$ for a target trace. 
As shown in Figure \ref{fig:method_max}, the CDF's highest magnitude of gradient change with a transition of $T_{intt}$ can be obtained by identifying the utmost outlier on the corresponding PDF. Algorithm \ref{algo:method_steepness} outlines how to examine the steepness of the curve of the target CDF through the corresponding PDF. It first calculates the PDF of $T_{intt}$ (cf. lines 1 $\sim$ 3). After that, the algorithm finds the best-fitting straight line through a set of $T_{intt}$ by using linear least squares regression analysis (cf. lines 4 $\sim$ 6).
In this algorithm, if there is $T_{intt}$, which is far from the best fitting straight line by more than a margin, we refer to $T_{intt}$ as an outlier. Note that, as the margin increases, the number of outliers decreases. As the final goal of this PDF analysis is to find the utmost outlier, we setup the margin at half the variance. This PDF analysis algorithm visits all $T_{intt}$ and collects the outliers for all categorized I/O instruction sets described in Section \ref{sec:method} (e.g., read/write and request sizes). Among the outliers, it first looks for the $T_{intt}$ with the maximum value (denoted by $T^{utmost}_{intt}$) and returns the distance, which is the difference between the f(x) value of the straight line at the utmost outlier and $PDF(T^{utmost}_{intt})$ (cf. line 14). Then, it compares the distances observed by each $PDF(T_{intt})$ and generates two graphs that have the top two highest $T^{utmost}_{intt}$ values (cf. line 15). 

\begin{figure}
\centering
\includegraphics[width=1\columnwidth]{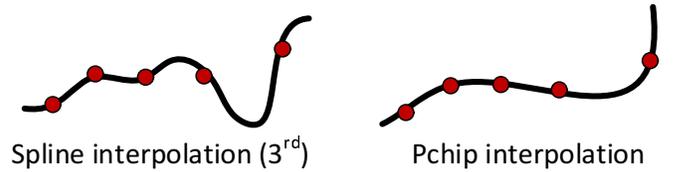}
\vspace{-15pt}
\caption{Different types of interpolations that we tested.\vspace{-8pt}}
\label{fig:method_interpolation}
\end{figure}

\noindent \textbf{Steepness analysis.}
It is a challenge to find the highest gradient change with a transition of $T_{intt}$ by analyzing a group of I/O requests with their CDF. Since CDF of $T_{intt}$ is a kind of non-differentiable function due to its discontinuity, the two I/O instruction groups selected by aforementioned graph classification algorithm must convert the discrete results into continuous results. While one can perform a curve fitting on $CDF(T_{intt})$ for the two groups to achieve a differentiable function, there is no perfect function that represents all variances observed in $CDF(T_{intt})$.
To address challenge, we interpolate $CDF(T_{intt})$ with piecewise nonlinear curve fitting; two interpolation methods are widely used: i) a special type of piecewise polynomial (called \emph{spline}) interpolation and ii) a piecewise cubic hermite interpolating polynomial (called \emph{pchip}) interpolation. As shown in Figure \ref{fig:method_interpolation}, spline evaluates the coefficient for each interval of data and has two continuous derivatives, whereas pchip has just one derivative, which preserves shape more smoothly than spline. Among $CDF(T_{intt})$ for all old block traces we tested, pchip exhibits the desired appearance of smoothness without oscillation and under/overfitting issues that spline has. 
Once we interpolate $CDF(T_{intt})$ with pchip, we can differentiate the results of interpolation and find the maximum of the differential, which is the highest magnitude of gradient change with a transition of $T_{intt}$. Note that, the analysis of $CDF(T_{movd})$ described earlier can be processed by the same curve fitting and differential calculation methods applied to $CDF(T_{intt})$.

\noindent \textbf{Hardware emulation and post-processing.}
Once the relative time costs are estimated, we can derive the $T_{sdev}$' equation, which infers the different device times under the execution of sequential reads/writes and random reads/writes.
In cases where there exist $n$ numbers of I/O instruction traced in the target workload, we can denote the idle time, inter-arrival time and device latency of the $i^{th}$ instruction (where $0 < i \leq n$) as $T^i_{idle}$, $T^i_{intt}$, and  $T^i_{sdev}$, respectively. We then visit each I/O instruction of an old trace and perform the following trace reconstruction procedure. First, we check the operation type and request size of the old trace's instruction and estimate $T^i_{sdev}$ using the $T_{sdev}$ model (cf. Section \ref{sec:method}). We also calculate $T^i_{intt}$ by checking the difference between time stamps of the $i^{th}$ and $i+1^{th}$ instructions, which are given by the old block trace. Thus, $T^i_{idle}$ exists if $T^i_{intt}$ is greater than $T^i_{sdev}$ (e.g., $T^i_{idle} = T^i_{intt} - T^i_{sdev}$). We then delay $T^i_{idle}$ using \texttt{sleep()} and issue the $i^{th}$ I/O instruction (composed of the same information of the old block trace) to the underlying brand-new device. We iterate this process for all $n$ I/O instructions. During this phase, we collect the new block trace using \emph{blktrace}, which is a standard block trace tool in Linux \cite{brunelle2006block}.
While this hardware emulation mimics the user behaviors, including system delays and idle periods, and incorporates actual channel delays and device times on the real target system, it is not feasible to inject synchronous/asynchronous mode information to each I/O request. Thus, we check the old trace and record all the indices of the instruction whose $T^i_{intt}$ is shorter than $T^i_{sdev}$. We then examine all the instructions of the new trace (but yet intermittent). In this post-processing, we subtract the new device time (measured by blktrace) from the corresponding inter-arrival time and update the next instruction based on the results, if the index of the instruction we are examining is in the range of instruction indices extracted by the old block trace. Note that if workloads provide the $T_{sdev}$ information, we can skip the $T_{sdev}$ inference phase, and immediately perform the hardware emulation and post processing after finding the short $T_{intt}$.


\section{Experimental Results}
\label{sec:resLults}
\begin{table*}[th]
\centering
\resizebox{\textwidth}{!}{%
\begin{tabular}{|c|c|c|c|c|c|c|c|c|c|c|c|c|c|c|c|c|}
\hline
Workload sets      & \multicolumn{8}{c|}{Microsoft Production Server (MSPS)}                               & \multicolumn{8}{c|}{FIU SRCMap}                                               \\ \hline
Published year     & \multicolumn{8}{c|}{2007}                                                             & \multicolumn{8}{c|}{2008}                                                     \\ \hline
Workloads          & 24HR           & 24HRS           & BS        & CFS   & DADS  & DAP    & DDR   & MSNFS & ikki   & madmax & online & topgun & webmail & casa   & webresearch & webusers \\ \hline
\# of block traces & 18             & 18              & 96        & 36    & 48    & 48     & 24    & 36    & 20     & 20     & 20     & 20     & 20      & 20     & 28          & 28       \\ \hline
Avg data size (KB) & 8.27           & 28.79           & 20.73     & 9.71  & 28.66 & 74.42  & 24.78 & 10.71 & 4.64   & 4.11   & 4.00   & 3.87   & 4.00    & 4.04   & 4.00        & 4.20     \\ \hline
Total size (GB)    & 21.2           & 178.6           & 331.2     & 43.6  & 44.6  & 84     & 44    & 317.9 & 25.4   & 3.8    & 22.8   & 9.4    & 31.2    & 80.4   & 13.7        & 33.6     \\ \hline
Workload sets      & \multicolumn{3}{c|}{FIU IODedup}             & \multicolumn{13}{c|}{MSR Cambridge (MSRC)}                                                                             \\ \hline
Published year     & \multicolumn{3}{c|}{2009}                    & \multicolumn{13}{c|}{2008}                                                                                             \\ \hline
Workloads          & \multicolumn{2}{c|}{mail+online} & homes     & mds   & prn   & proj   & prxy  & rsrch & src1   & src2   & stg    & web    & wdev    & usr    & hm          & ts       \\ \hline
\# of block traces & \multicolumn{2}{c|}{21}          & 21        & 2     & 2     & 5      & 2     & 3     & 3      & 3      & 2      & 4      & 4       & 3      & 1           & 1        \\ \hline
Avg data size (KB) & \multicolumn{2}{c|}{4.0}         & 5.23      & 33.0  & 15.4  & 29.6   & 8.6   & 8.4   & 35.7   & 40.9   & 26.2   & 7      & 34      & 38.65  & 15.16       & 9.0      \\ \hline
Total size (GB)    & \multicolumn{2}{c|}{57.1}        & 84.6      & 208.4 & 568.8 & 4780.1 & 4353  & 27.63 & 6516.5 & 230.6  & 226.4  & 625.4  & 23.7    & 5506.1 & 9.24        & 16.2     \\ \hline
\end{tabular}}
\caption{Important characteristics of the publicly-available conventional block traces that we reconstructed.\vspace{-5pt}}
\label{tab:Workload Characteristics}
\end{table*}

In this evaluation, we focus on answering the following questions: i) How accurate is our inference model? ii) How realistic can our hardware/software make $T_{intt}$ compared to conventional approaches? and iii) What are the system implications based on the revisions of $T_{idle}$?

\noindent \textbf{Evaluation node.} For the target system where we reconstruct block traces, we build up a storage node that employs an all-flash array by grouping four NVM Express SSDs \cite{NVMeSSD}. The storage capacity of each SSD is 400GB, and a single device consists of 18 channels, 36 dies, and 72 planes. Our storage node can exhibit different levels of parallelism, ranging from an array to channel, channel to die, which in turn can offer read and write bandwidths as much as 9GB/s and 4GB/s, respectively. Our all-flash array is connected to the node's north-bridge via four PCIe 3.0 slots, each containing four lanes \cite{ref-if-pcie3} to the storage node. 

\noindent \textbf{Target block traces.} We reconstruct three workload categories: i) Florida International University (FIU) \cite{verma2010srcmap, koller2010deduplication}, ii) Microsoft Production Server (MSPS) \cite{kavalanekar2008characterization}, and iii) Microsoft Research Cambridge (MSRC) \cite{narayanan2008write}. Together, FIU, MSPS, and MSRC contain a total of 577 block traces, which are used for a wide spectrum of simulation-based studies \cite{zhang2013warming, jeong2014lifetime, narayanan2009migrating, soundararajan2010extending}.
FIU workloads offer university-scale production server characteristics, which consist of two different types of sub-workloads: SRCMap and IODedup. While SRCMap workloads are collected for an application that optimizes system energy by virtualizing storage, department-level virtual machines for web services and mail, file and version control servers are collected by IODedup. On the other hand, MSPS provides eight different kinds of production server scenarios, and MSRC provides thirteen kinds of data center server scenarios.
In MSRC, all workloads contain specific device-level information such as the type of RAID, while the same information for most workloads in MSPS is unknown. In addition, MSPS and MSRC workloads are collected by using an event-based kernel-level tracing facility \cite{park2003server} which can capture detailed information such as issue and completion time stamps; These timestamps are captured when requests are issued from a device driver to the target disk and when the disk completes the I/O operations, respectively.
Note that even though all the traces related to the three categories of workloads discussed above include various system configurations and have a wide range of user scenarios, they are all collected around 2007$\sim$2009 on disk-based systems. The important characteristics of traces, including the size, and the number of traces per workload, are listed in Table \ref{tab:Workload Characteristics}.
  
\noindent \textbf{Reconstruction techniques.} 
We evaluate five different block reconstruction methods:

\begin{itemize}[leftmargin=8pt, itemsep=-1ex,topsep=-1ex,partopsep=0ex,parsep=1ex]
\item \texttt{Acceleration}: Reconstruction by shortening $T_{intt}$ \cite{jeong2014lifetime}.
\item \texttt{Revision}: Replaying block traces on all-flash array \cite{chen2011tpc}.
\item \texttt{Fixed-th}: An advanced revision method by inferring $T_{idle}$ with a fixed threshold.
\item \texttt{Dynamic}: Reconstructions using our inference model, but with no post-processing.
\item \texttt{TraceTracker}: Hardware/software co-evaluation for trace reconstruction.
\end{itemize}

We leverage the value (i.e., 100) that a simulation-based SSD work uses \cite{jeong2014lifetime} for its $T_{intt}$ acceleration. On the other hand,  \texttt{Fixed-th} considers the worst-case device latency of old storage with a fixed threshold value and uses it for inferring $T_{idle}$. To select a reasonable threshold, we performed a different set of evaluations on a HDD-based node with various thresholds, ranging from 10 ms to 100 ms, and selected 10 ms as \texttt{Fixed-th}'s optimal threshold. In contrast, \texttt{Dynamic} injects different $T_{idle}$ per I/O instruction by speculating it over our inference model, but without the post-processing component of \texttt{TraceTracker}.


\begin{figure}
\centering
\begin{subfigure}{0.45\columnwidth}
\includegraphics[width=\columnwidth]{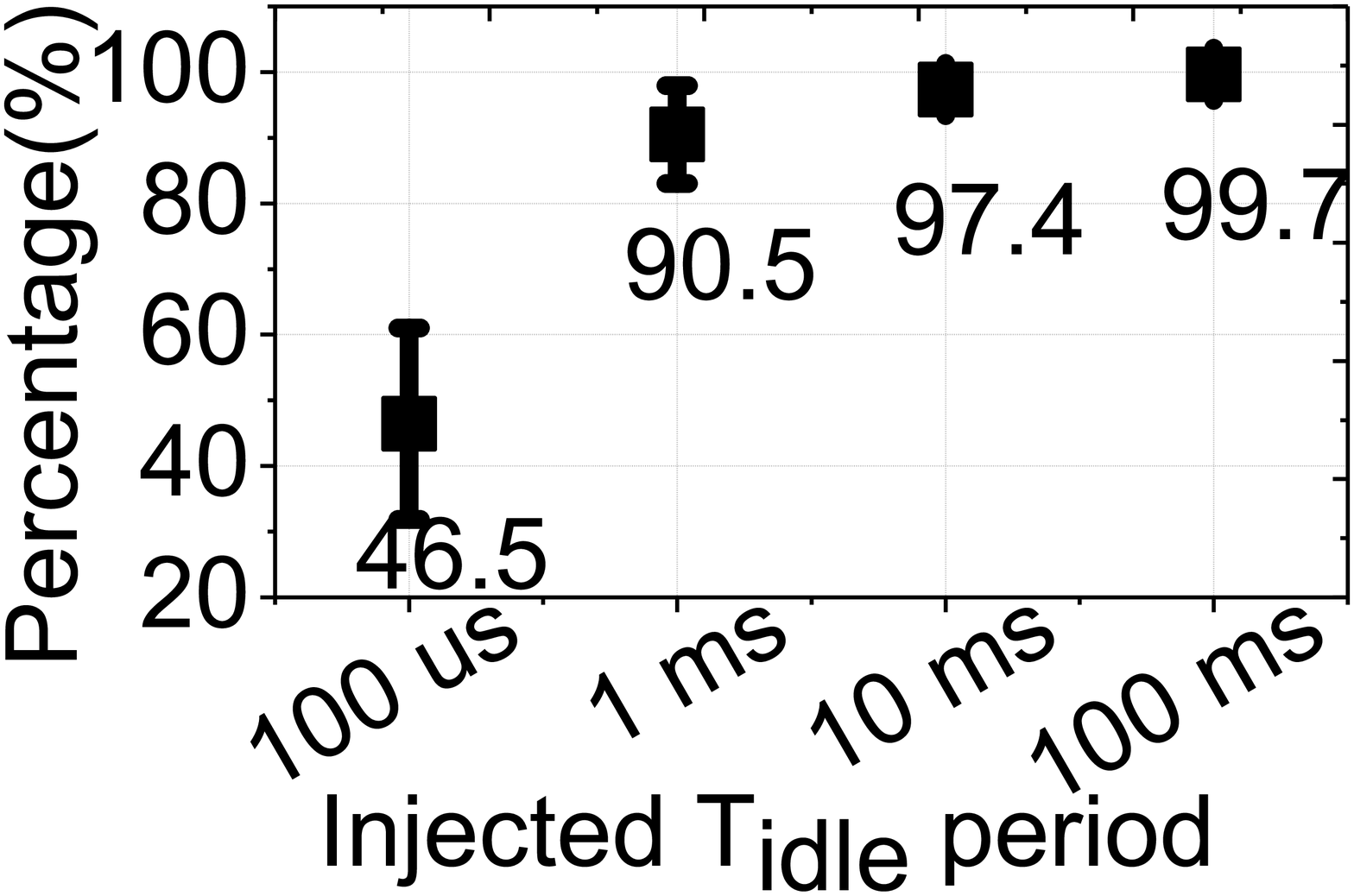}
\caption{$T_{sdev}$ known traces.}
\label{fig:eval_verif1_TP}
\end{subfigure}
~
\begin{subfigure}{0.45\columnwidth}
\includegraphics[width=\columnwidth]{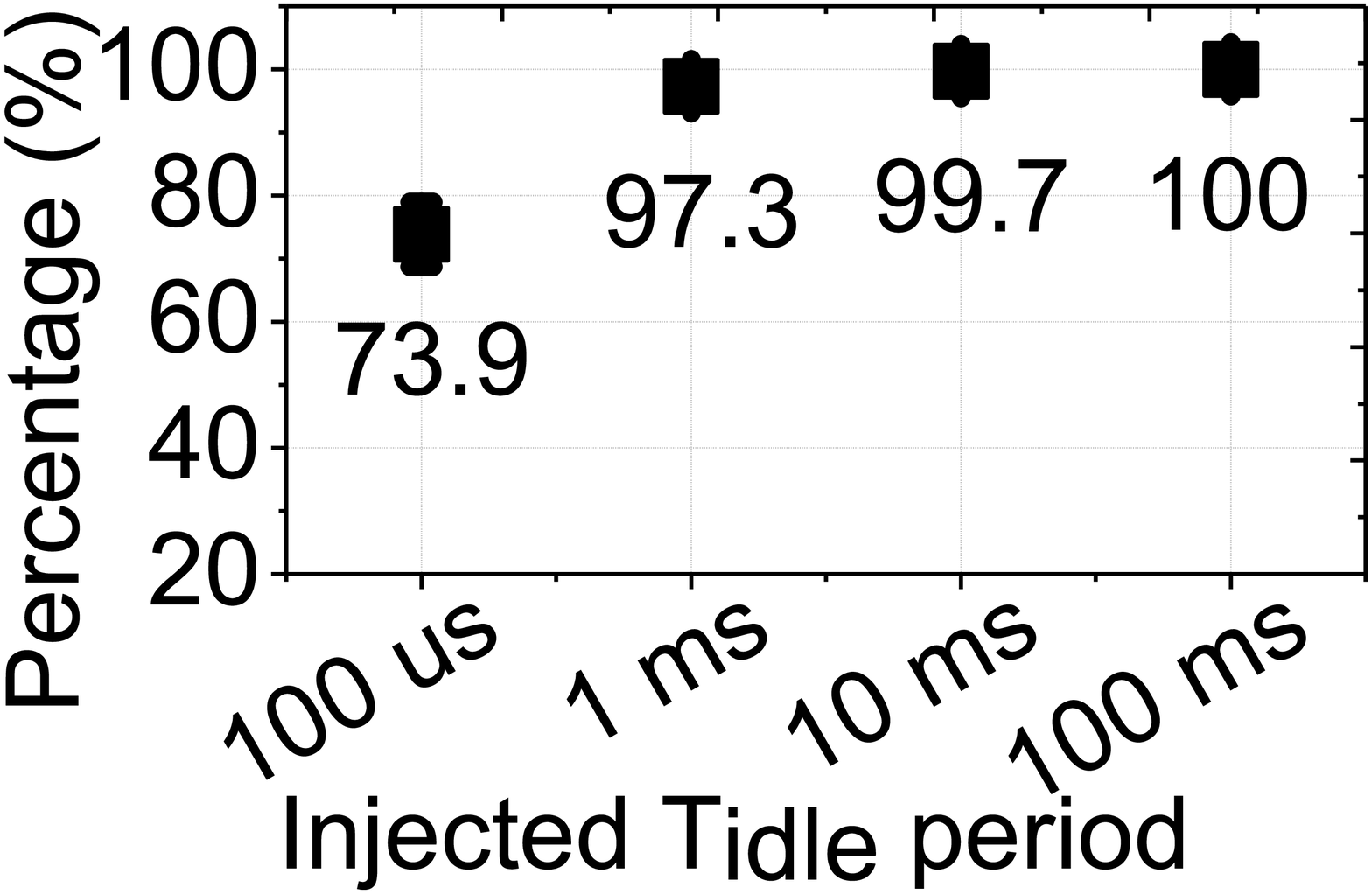}
\caption{$T_{sdev}$ unknown traces.}
\label{fig:eval_verif2_TP}
\end{subfigure}
\vspace{8pt}
\caption{Verification results, Len(TP).}
\label{fig:eval_verif_TP}
\end{figure}

\begin{figure}
\centering
\begin{subfigure}{0.45\columnwidth}
\includegraphics[width=\columnwidth]{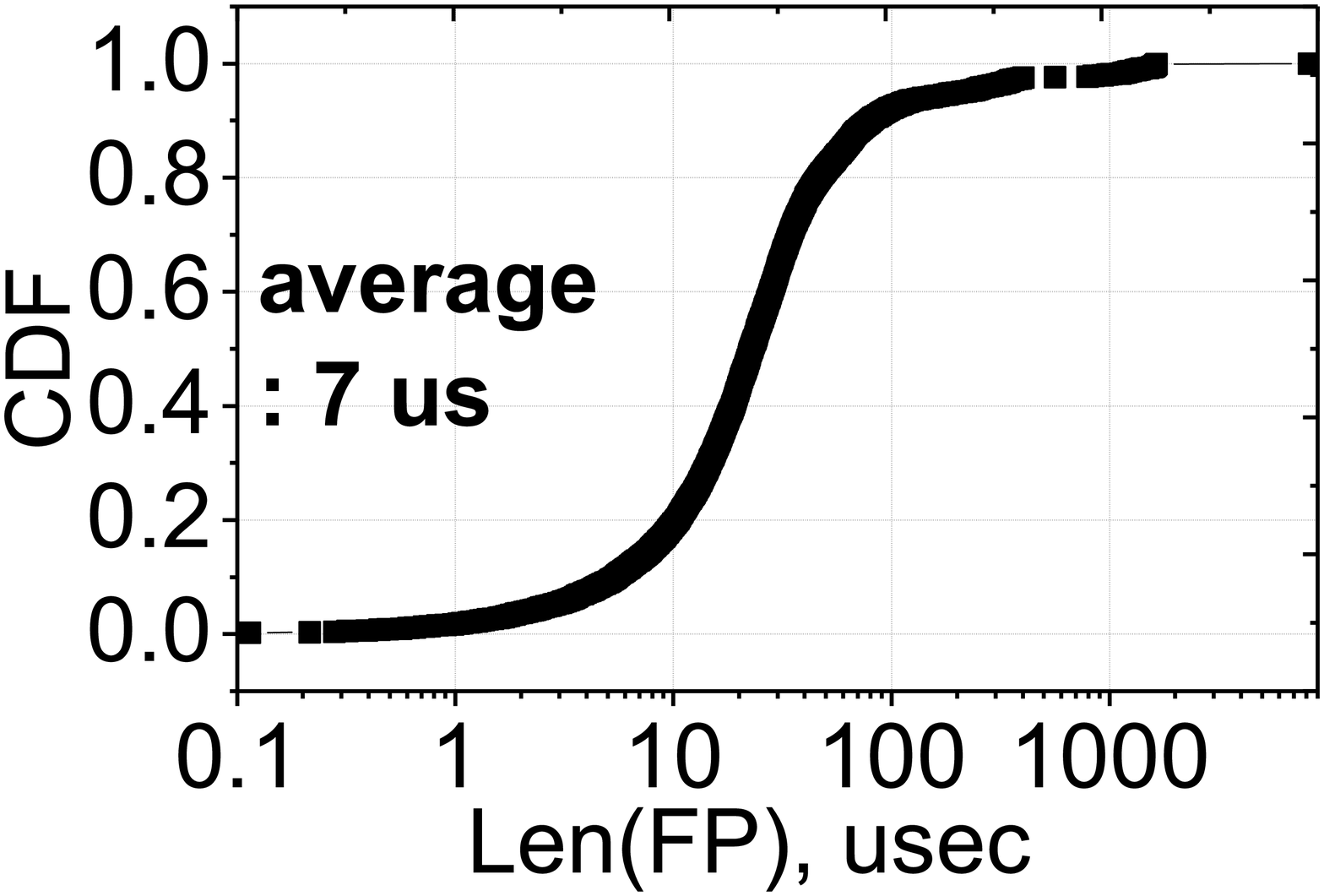}
\caption{$T_{sdev}$ known traces.}
\label{fig:eval_verif1_FP}
\end{subfigure}
~
\begin{subfigure}{0.45\columnwidth}
\includegraphics[width=\columnwidth]{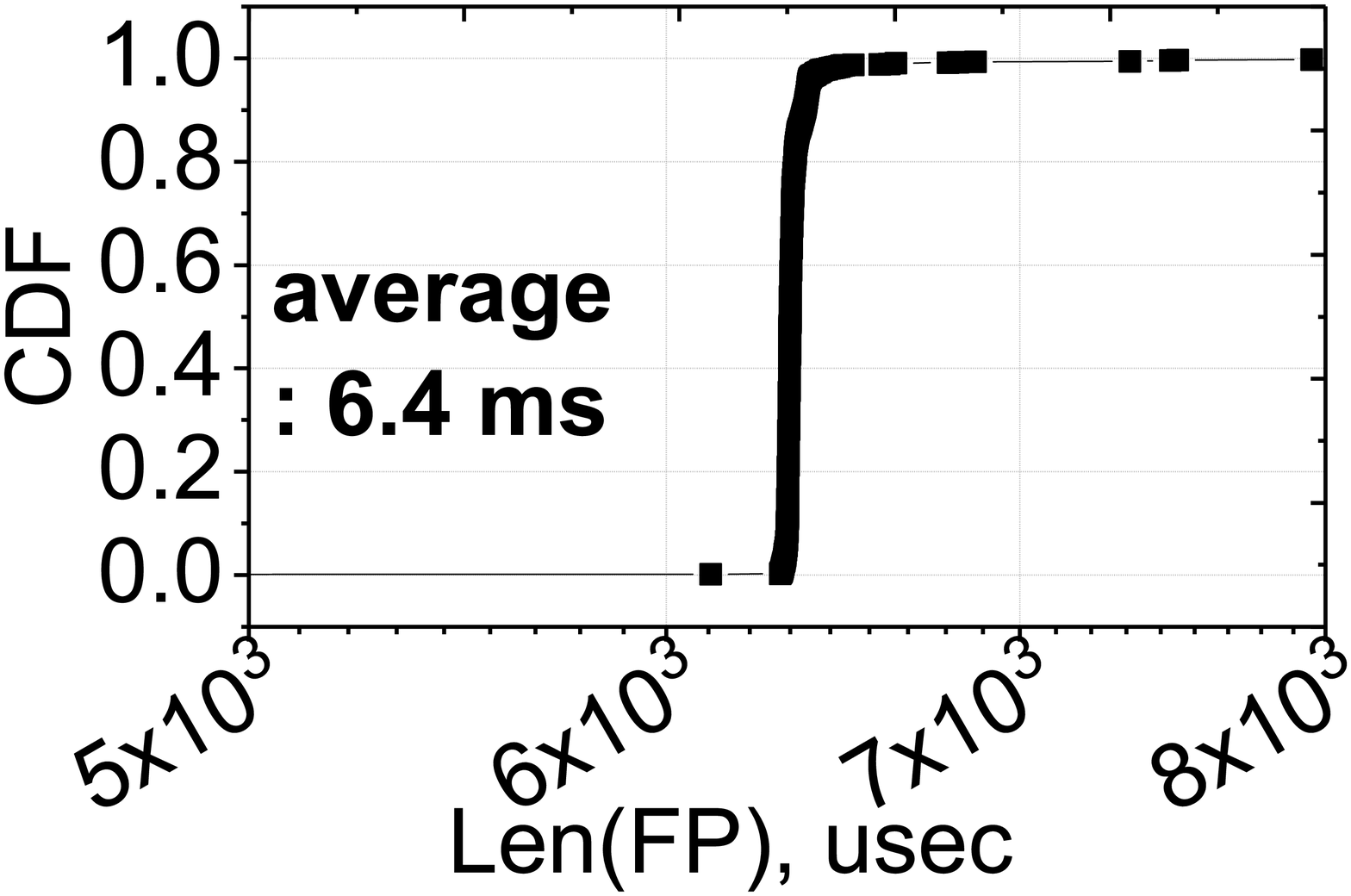}
\caption{$T_{sdev}$ unknown traces.}
\label{fig:eval_verif2_FP}
\end{subfigure}
\vspace{8pt}
\caption{Verification results, Len(FP).\vspace{-5pt}}
\label{fig:eval_verif_FP}
\end{figure}

\subsection{Verification}
\noindent \textbf{Metrics.} The results of this verification evaluation can be either positive or negative, each of which may be true or false. If the inference model speculates that there is $T_{idle}$, it can be classified as positive, and otherwise, the result is negative. Being negative or positive can be tested per I/O instruction. On the other hand, if the existence of $T_{idle}$ is same in both target and reconstructed block traces, one can call this as true. Otherwise, it is false.
Therefore, the results of the inference model test can be represented by four different statistics: true positive (TP), false positive (FP), false negative (FN), and true negative (TN). For verification, we will use four functions as follows: i) $Detection(TP) =$ \emph{number of} $TP$ $/$ \emph{number of} $T^{injected}_{idle}$, ii) $Detection(FP) =$ \emph{number of} $FP$ $/$ \emph{total number of I/O instructions}, iii) $Len(TP) = T^{estimated}_{idle}$ $/$ $T^{injected}_{idle}$, and iv) $Len(FP) =  T^{estimated}_{idle}$, where $T^{injected}_{idle}$ and $T^{estimated}_{idle}$ are the idle times that were injected into the target block traces and speculated by our inference model, respectively. The first two functions capture the ratios of the number of TP/FP and the number of corresponding I/O instructions, whereas $Len(TP)$ and $Len(FP)$ indicate how much our inference model speculates accurate $T_{idle}$ based on the result of a prediction hit or miss, respectively. Note that $Len(TP)$ is the ratio of the speculated idle periods and actual idle periods, whereas $Len(FP)$ is the actual period that the inference model mispredicts.  

\noindent \textbf{Results.} 
Since the block traces have no information on $T_{idle}$, we inject $T_{idle}$ in random places with various idle periods, ranging from 100 us to 100 ms. In this evaluation, injected $T_{idle}$ accounts for 10\% of the total I/O instructions of the target block traces. We compare the injected $T_{idle}$ with the $T_{idle}$ predicted by our inference model. We select two different groups of traces. One includes the traces that contain no timing information (e.g., FIU), and the other has I/O submission and completion time information (e.g., MSPS), which can be considered as $T_{sdev}$. In this evaluation, we denote the former and the latter as $T_{sdev}$ known traces and $T_{sdev}$ unknown traces, respectively.  

Figure \ref{fig:eval_verif_TP} shows the results of $Len(TP)$ observed by two trace groups that \texttt{TraceTracker} reconstructed. If the injected $T_{idle}$ is longer than 1 ms, \texttt{TraceTracker} shows 90.5\% and 97.3\% accuracy of TP for $T_{sdev}$ known traces and $T_{sdev}$ unknown traces, respectively. If the injected $T_{idle}$ is close to 100 us, the accuracy of TP declines compared to other cases by 46.5\% and 73.9\% for $T_{sdev}$ known traces and $T_{sdev}$ unknown traces, respectively. This is because the injected $T_{idle}$ is in a range of the latency that new storage (in our case, Intel NVMe 750) exhibits. While this blurring boundary issue could make it difficult for our inference model to distinguish between device latency and idle time, most of the actual microsecond-scale system delays and idle periods are revived by our inference model. In addition, we observed that the results of $Detection(TP)$ is in the range of 82.2\% $\sim$ 99.7\% across all the block traces that \texttt{TraceTracker} built. 
On the other hand, $Detection(FP)$ is, on average, 6\% and 26\% while $Len(FP)$ is on average 7 us and 6.4 ms for $T_{sdev}$ known traces and $T_{sdev}$ unknown traces, respectively. However, the distribution of $Len(FP)$ observed by the reconstructed traces tells a different story. As shown in Figure \ref{fig:eval_verif_FP}, more than 98\% of $Len{FP}$ for $T_{sdev}$ known traces and $T_{sdev}$ unknown traces are shorter than 1 ms and 6 ms, respectively. Considering the high accuracy of TP and low impact of FP, we can conclude that \texttt{TraceTracker} is within the confidence interval to reconstruct $T_{intt}$.

\begin{figure}
\centering
\begin{subfigure}{0.45\columnwidth}
\includegraphics[width=\columnwidth]{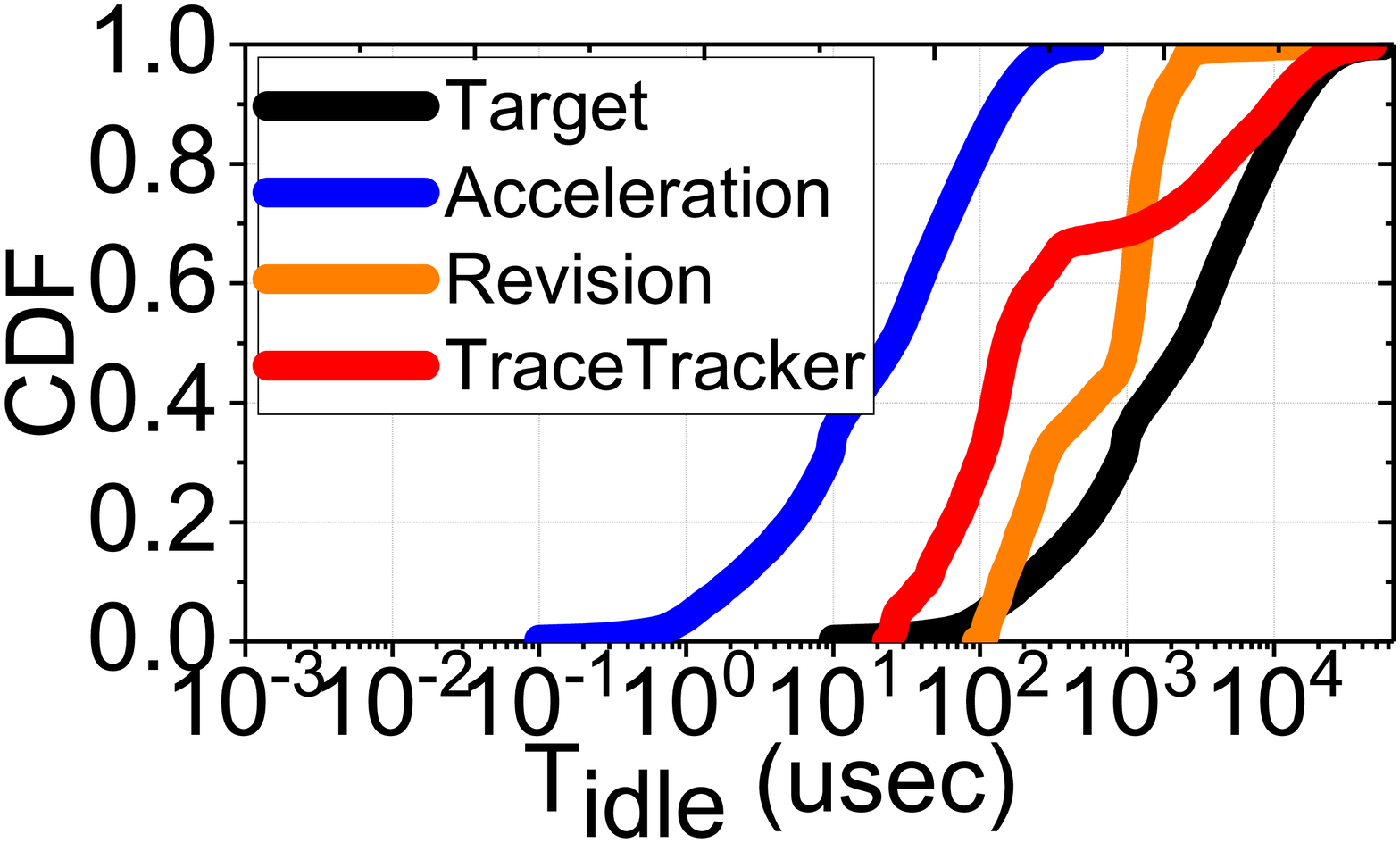}
\caption{Unaware of $T_{idle}$.}
\label{fig:eval_case_all1}
\end{subfigure}
~
\begin{subfigure}{0.45\columnwidth}
\includegraphics[width=\columnwidth]{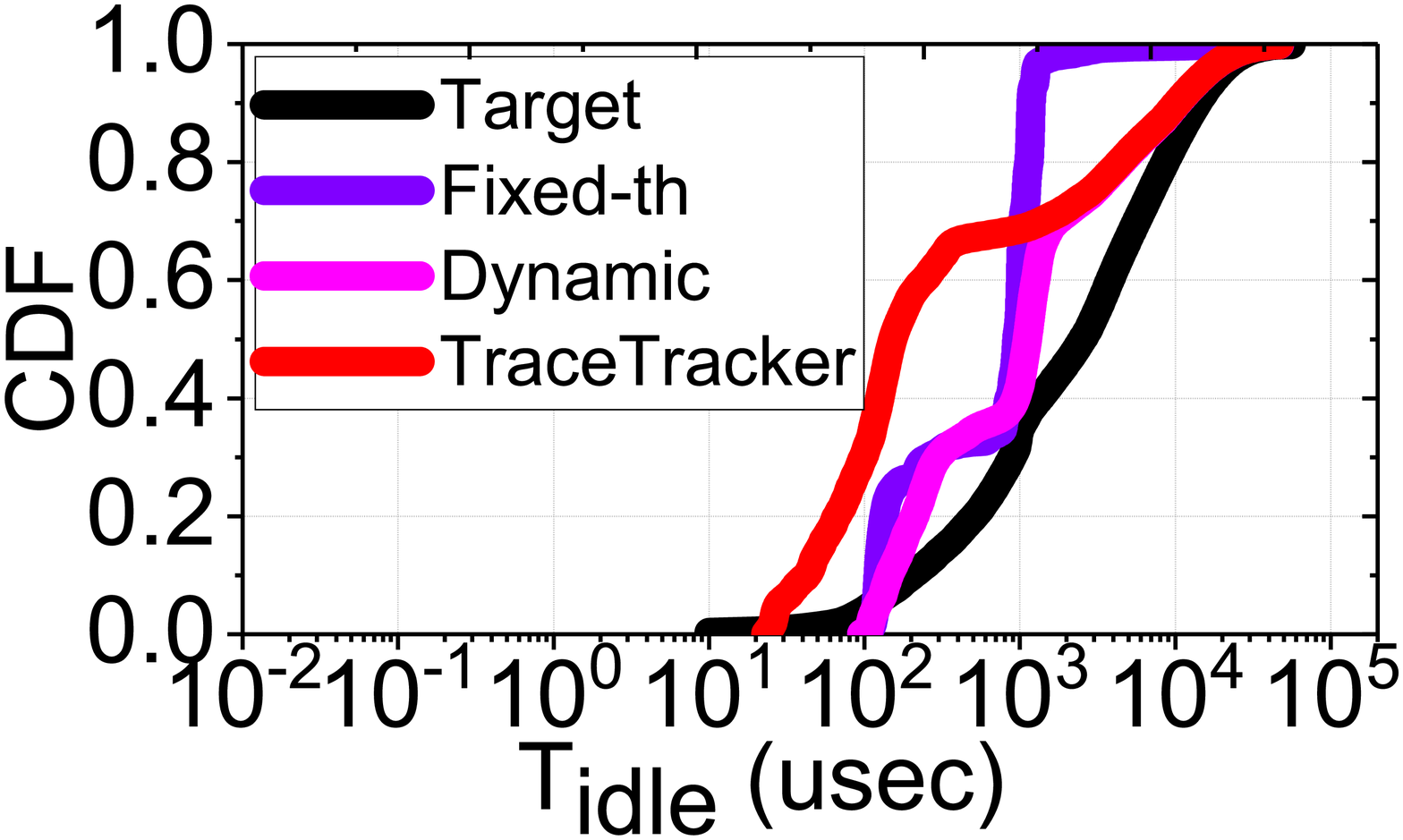}
\caption{Aware of $T_{idle}$.}
\label{fig:eval_case_all2}
\end{subfigure}
\vspace{8pt}
\caption{CDF distribution of $T_{intt}$ (\textit{MSNFS}).}
\label{fig:eval_case_all}
\end{figure}

\begin{figure}
\centering
\includegraphics[width=1\columnwidth]{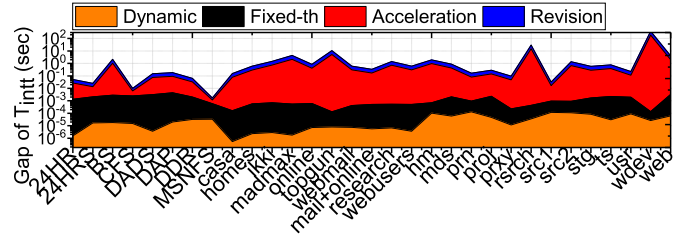}
\caption{$T_{intt}$ differences among the different kinds of trace reconstruction techniques and \texttt{TraceTracker} method.\vspace{-5pt}}
\label{fig:eval_case_gapScheme}
\end{figure}

\noindent \textbf{Comparisons.}
In this section, we analyze the accuracy of \texttt{TraceTracker} compared to other reconstruction methods by inspecting the details of the $T_{intt}$. To this end, we compare \texttt{TraceTracker}'s CDF of $T_{intt}$ with two different groups of methods, each being unaware of and aware of $T_{idle}$; the results are shown in Figures \ref{fig:eval_case_all1} and \ref{fig:eval_case_all2}, respectively.
In these figures, \texttt{Target} shows, the CDF of $T_{intt}$ brought by the original block traces collected on HDD-based nodes. One can observe from Figure \ref{fig:eval_case_all1} that, \texttt{Acceleration} just shifts the CDF of \texttt{Target} from the right to the left as much as the acceleration factor indicates (e.g., 100x), which eliminates all the useful information to simulate target systems. On the other hand, \texttt{Revision} reflects the characteristics of the underlying new storage. However, compared to \texttt{TraceTracker}, it removes $T_{cdel}$ and  $T_{idle}$ by around 70\% and 30\%, respectively. As shown in Figure \ref{fig:eval_case_all2}, while \texttt{Fixed-th} and \texttt{Dynamic} behave more realistically than \texttt{Acceleration}, unfortunately, \texttt{Fixed-th} loses 65\% of the $T_{idle}$ and \texttt{Dynamic} exhibits 30\% longer $T_{intt}$ than \texttt{TraceTracker} as it also loses asynchronous/synchronous mode information and is unable to capture $T_{cdel}$ appropriately. 

Figure \ref{fig:eval_case_gapScheme} plots the average difference between \texttt{TraceTracker} and other trace reconstruction methods in terms of $T_{intt}$ for all the workloads we tested. One can observe from the figure that \texttt{Acceleration} and \texttt{Revision} methods that possess no information for $T_{idle}$ to reconstruct traces differ by 7.08 and 7.15 seconds from \texttt{TraceTracker}, respectively. Considering the worst-case latency of the underlying SSD accesses (around 2 ms), losing such idle times that include system delays and user behaviors can have a great impact on diverse simulation-based studies. While \texttt{Fixed-th} and \texttt{Dynamic} have less $T_{intt}$ differences compared to \texttt{Acceleration} or \texttt{Revision}, the difference between \texttt{TraceTracker} and their $T_{intt}$ is as high as 1.3 ms and 0.035 ms, respectively.
This means that, even though  \texttt{Fixed-th} and \texttt{Dynamic} can capture the underlying storage characteristics, the actual time behaviors, including $T_{cdel}$ and  $T_{idle}$, are omitted. As a result, they can exhibit different system behaviors with inaccurate $T_{intt}$ values.

\subsection{System implications}

\begin{figure}
\centering
\includegraphics[width=1\columnwidth]{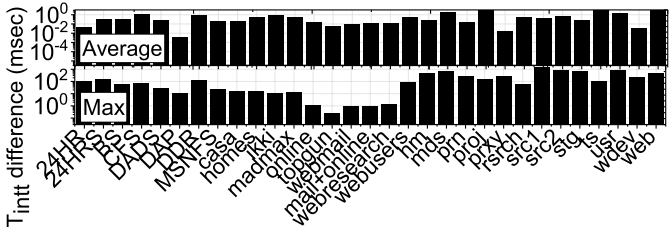}
\caption{$T_{intt}$ differences between the target block traces and \texttt{TraceTracker} traces.}
\label{fig:eval_case_gap}
\end{figure}

\begin{figure}
\begin{subfigure}{0.45\columnwidth}
\includegraphics[width=\columnwidth]{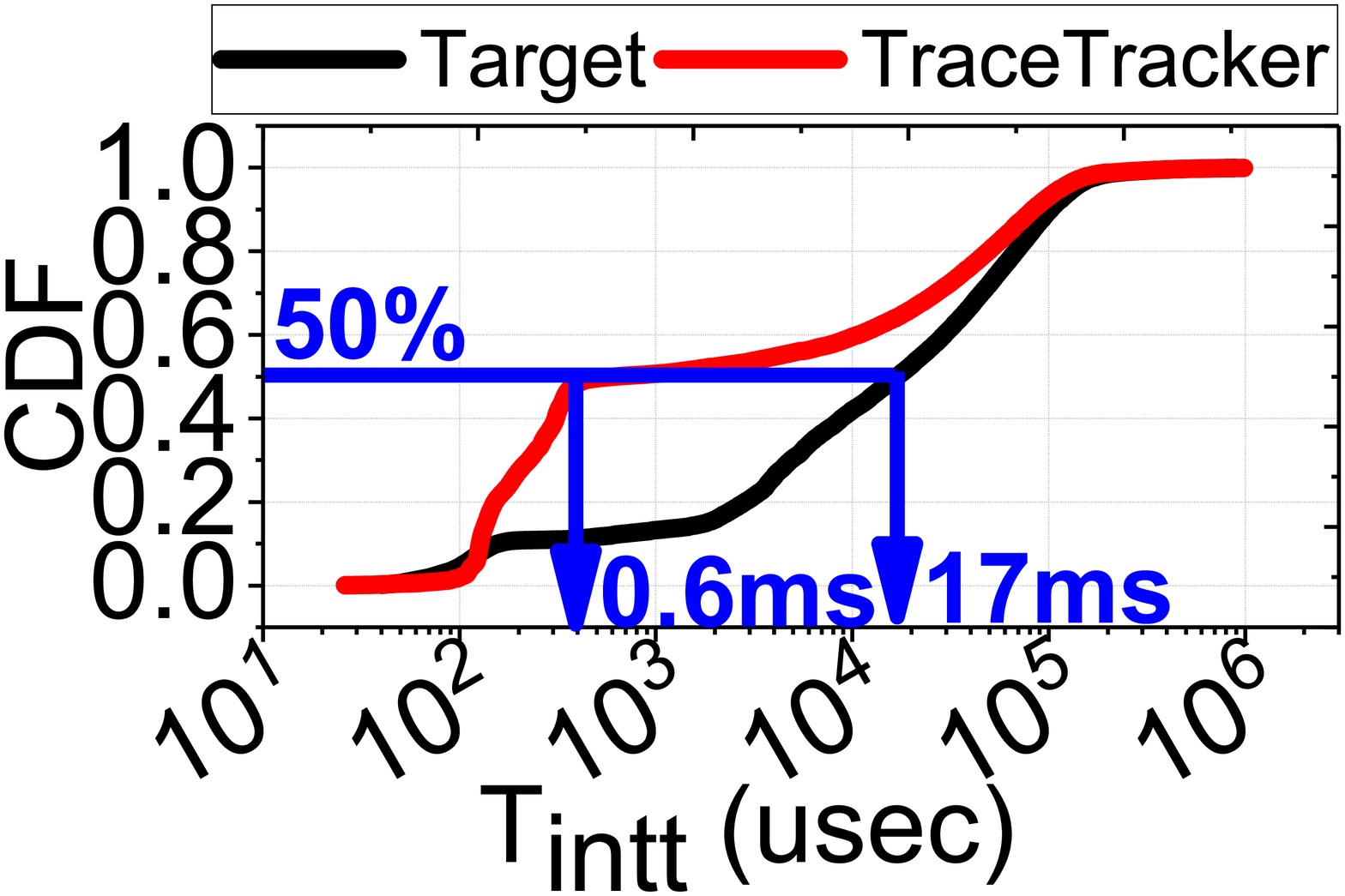}
\caption{\textit{CFS} (MSPS).}
\label{fig:eval_case_CDF_CFS}
\end{subfigure}
~
\begin{subfigure}{0.45\columnwidth}
\includegraphics[width=\columnwidth]{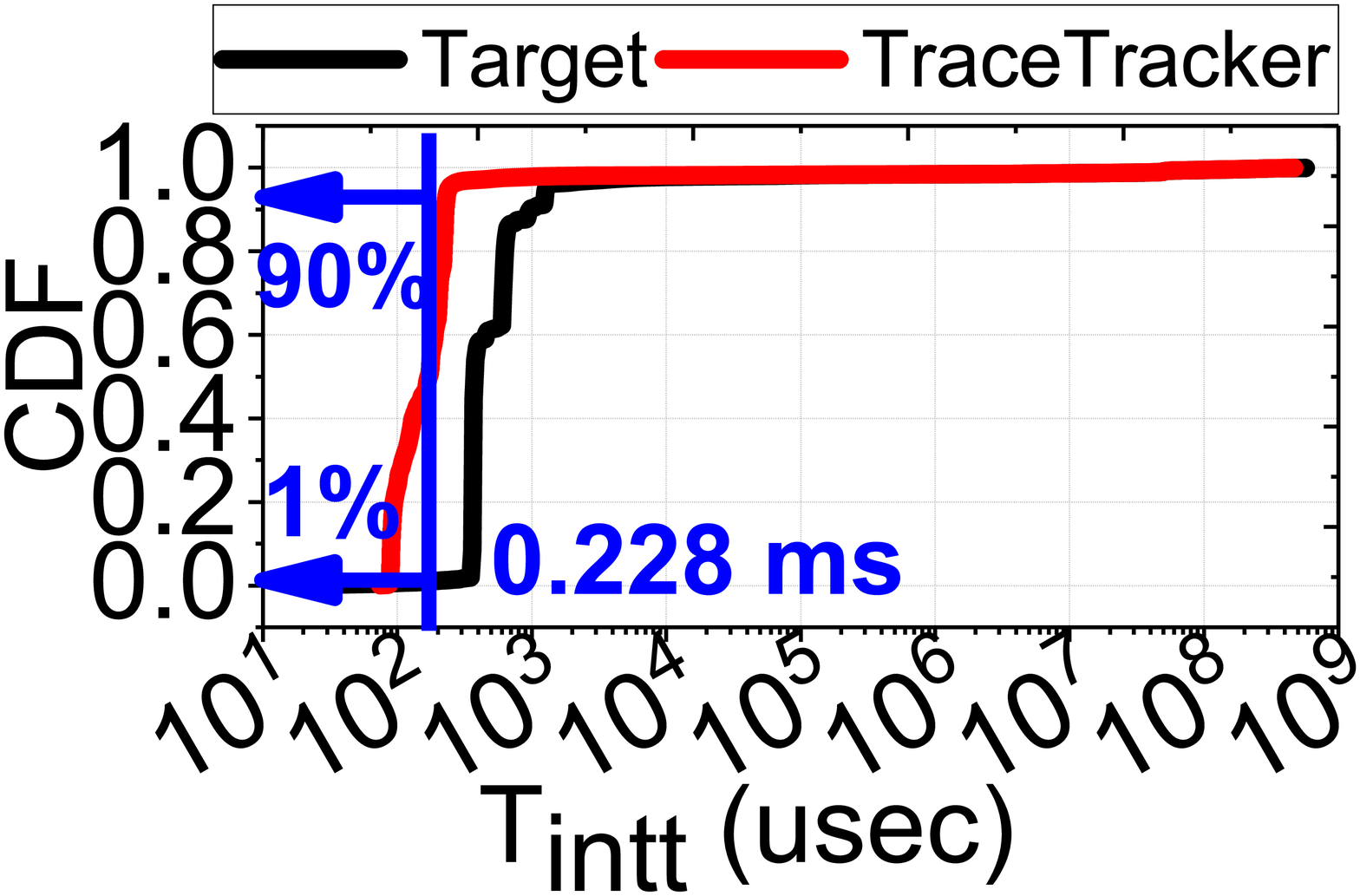}
\caption{\textit{ikki} (FIU).}
\label{fig:eval_case_CDF_ikki}
\end{subfigure}
\vspace{10pt}
\caption{Distribution differences between the target block traces and \texttt{TraceTracker} traces.}
\label{fig:eval_case_CDF}
\end{figure}

\noindent\textbf{Overall analysis of inter-arrival times.}
The top and bottom of Figure \ref{fig:eval_case_gap} plot the average and maximum $T_{intt}$ differences between the target block traces and traces reconstructed by \texttt{TraceTracker}. As shown in the figure, $T_{intt}$ of the \texttt{TraceTracker} traces is 0.677 ms shorter, on average, than that of the target block traces. The $T_{intt}$ implies that system analysis and evaluation studies that use the $T_{intt}$ of target block traces should consider the \texttt{TraceTracker} traces instead since the time budget to perform foreground/background tasks can be tightened when the storage system is changed. For example, the \textit{ts} workload (MSRC) has an average of 3 ms shorter $T_{intt}$ in \texttt{TraceTracker} traces than in the target block traces. In addition, the median values of $T_{intt}$ are 2 ms and 0.02 ms for target block traces and \texttt{TraceTracker}, respectively. Note that, the average $T_{intt}$ differs among the 31 workloads because of the impact of the specific workload characteristics such as request size and type.

To analyze $T_{intt}$ differences between the two traces in detail, we plot the CDF distribution of $T_{intt}$, as shown in Figure \ref{fig:eval_case_CDF}, only for the \textit{CFS} (MSPS) and \textit{ikki} (FIU), which have the maximum $T_{intt}$ differences among the same workload categories (MSPS, FIU). As shown in the figures, the $T_{intt}$ distribution of the \texttt{TraceTracker} traces leans towards the short time period and the average differences are 1 ms and 0.823 ms, respectively. For instance, 50 \% of $T_{intt}$ in the target block traces are less than 17 ms while that of the \texttt{TraceTracker} is 0.601 ms in Figure \ref{fig:eval_case_CDF_CFS}. In addition, as shown in Figure \ref{fig:eval_case_CDF_ikki}, 1 \% of $T_{intt}$ in the target block traces are less than 0.228 ms, while 90 \% of the $T_{intt}$ are less than the same value in the \texttt{TraceTracker} traces.

\begin{figure}
\centering
\includegraphics[width=1\columnwidth]{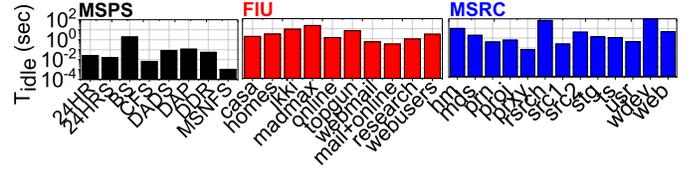}
\caption{Average time period of $T_{idle}$.\vspace{4pt}}
\label{fig:eval_UI_len}
\end{figure}

\begin{figure}
\centering
\includegraphics[width=1\columnwidth]{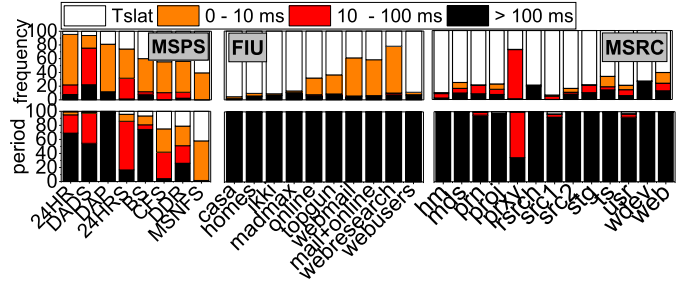}
\caption{Breakdown of $T_{idle}$.}
\label{fig:eval_break}
\end{figure}

\noindent\textbf{Details of idle times.}
$T_{idle}$ can be a representative workload characteristic, and the estimated $T_{idle}$ was injected when the traces are reconstructed on the target storage system. Since the $T_{idle}$ periods should be same in reconstructed traces, the $T_{idle}$ that we estimated can be immediately used for other conventional block traces. Figure \ref{fig:eval_UI_len} shows the $T_{idle}$ period estimated by \texttt{TraceTracker}. As shown in the figure, the average $T_{idle}$ of MSPS is 0.27 s, and that of FIU is 2.80 s remove \textit{madmax} workload has 20.5 s of longest $T_{idle}$ among the FIU workloads. MSRC has an average $T_{idle}$ value of 2.25 s, except for \textit{rsrch} and \textit{wdev} which have 69.2 s and 403.1 s of $T_{idle}$, respectively.

To check the detailed $T_{idle}$ patterns of the 31 workloads, we analyze the breakdown of total $T_{intt}$ duration by grouping these into $T_{slat}$, $T_{idle}$ (0 $\sim$
 10 ms), $T_{idle}$ (10 $\sim$
 100 ms), and $T_{idle}$ (longer than 100 ms). The top and bottom parts of Figure \ref{fig:eval_break} focus on the frequency and period, respectively. The frequency refers to the total number of requests per group while the period means the total time duration of each group.
As shown in the figure, the MSPS workloads have larger $T_{idle}$ breakdown in terms of frequency, compared to other workloads; the average $T_{idle}$ breakdown is 70\%, 31\%, and 26\% for MSPS, FIU, and MSRC, respectively. In contrast to the frequency, the average breakdown of period per workload categories is 87\%, 99.8\%, and 99.2\% for MSPS, FIU, and MSRC, respectively.
In other words, although the FIU and the MSRC workloads have low $T_{idle}$ frequency, most of the $T_{intt}$ is $T_{idle}$ at around 90\%. In addition, as shown in the figure, most of the $T_{idle}$ is longer than 100 ms in the FIU and MSRC workloads. Similar to the average $T_{idle}$ period shown in Figure \ref{fig:eval_UI_len}, the breakdown pattern of MSPS workloads varies compared to other workloads. In the MSPS, the average frequency is 30\%, 47.7\%, 15\%, and 6.7\% for each group, while the average period breakdown is 12.6\%, 18.3\%, 26\%, and 42.7\%, respectively. Since the MSPS workloads have short $T_{idle}$, it is harder for them to utilize inter-arrival times, compared to other workloads.

\section{Related Works}
\label{sec:related}
There exist many prior studies that proposed to modify the conventional block traces to adjust to new storage systems \cite{jeong2014lifetime, trushkowsky2011scads, weddle2007paraid, chen2011tpc, ganger1998using, mesnier2007trace, zhu2005tbbt}. For example, \cite{jeong2014lifetime, trushkowsky2011scads, weddle2007paraid} tried to simply accelerate the inter-arrival times with a fixed scaling factor. On the other hand, \cite{chen2011tpc, ganger1998using, mesnier2007trace, zhu2005tbbt} replayed I/O requests on the real storage system by injecting an extra delay or zero (no-idle) between two consecutive requests. In contrast, \cite{mesnier2007trace} was aware of the behaviors of parallel applications, which are widely used in scientific or business environments, and reflected these onto target block traces by injecting different idle times per I/O instruction. As there are multiple nodes that execute parallel applications, this work calculates the duration of an extra delay by taking into account the computing time and synchronization time which is required for each node to ensure data synchronization; the input data of one node is another node's output. While all the above methods did not consider user behaviors and I/O execution mode, \texttt{TraceTracker} can reconstruct the old block traces irrespective of application types and can classify the inter-arrival times into I/O subsystem latency and extra delay (idle times) by including both system and user behaviors. In \texttt{TraceTracker}, the idle times are decided by modeling the performance of the target trace's storage system.

Early studies on storage performance modeling \cite{merchant1996analytic, mesnier2006relative, shriver1998analytic, uysal2001modular, wang2004storage} try to capture the performance of new storage systems by identifying the target workload's characteristics. For example, \cite{wang2004storage} used Classification and Regression Trees (CART), which is learning-based black box modeling technique. However, CART does not understand the input features and generates a multidimensional function of the model. Thus, for the storage performance, \cite{wang2004storage} utilized the request information (e.g., inter-arrival times, logical block number, request type, and data size) as features of the CART algorithm. Unfortunately, the main problem of machine-learning based modeling is that it is hard to explain how the model can be achieved. While \cite{wang2004storage} creates performance model without understanding the inter-arrival times, our \texttt{TraceTracker} analytically models storage performance by decomposing the inter-arrival times and detecting the short inter-arrival times for asynchronous I/O execution.

\section{Acknowledgement}
\label{sec:ack}
This research is mainly supported by NRF 2016R1C1B2015312. This work is also
supported in part by DOE DE-AC02-05CH 11231, IITP-2017-2017-0-01015,
NRF-2015M3C4A7065645, and MemRay grant (2015-11-1731). Kandemir is supported
in part by NSF grants 1439021, 1439057, 1409095, 1626251, 1629915, 1629129
and 1526750, and a grant from Intel. Myoungsoo Jung is the corresponding
author.

\section{Conclusion}
\label{sec:conclusion}
\texttt{TraceTracker} is a new approach to reconstruct existing block traces to new traces which is being aware of the target storage system only with inter-arrival time information of target workloads. To maintain the important workload's characteristics such as system and user behaviors in the new traces, \texttt{TraceTracker} estimates the idle times by automatically inferencing the performance of storage system from target block traces. We can detect 99\% of system delays and idle periods appropriately and secure the corresponding idle periods by 96\% of a real execution, on average.

\end{document}